\documentclass[11pt]{article}
\usepackage[margin=1in]{geometry}
\usepackage{amsmath}
\usepackage[T1]{fontenc}
\usepackage[utf8]{inputenc}
\usepackage{libertine}
\usepackage[libertine]{newtxmath}
\usepackage{microtype}
\usepackage{graphicx}
\usepackage{booktabs}
\usepackage{threeparttable}
\usepackage[font=small,labelfont=bf]{caption}
\usepackage[round,authoryear,sort]{natbib}
\makeatletter
\@ifundefined{ifNAT@swa}
  {\DeclareRobustCommand{\harvardand}{and}}
  {\DeclareRobustCommand{\harvardand}{\ifNAT@swa\&\else and\fi}}
\makeatother
\usepackage{etoolbox}
\usepackage{xcolor}
\definecolor{linkcolor}{HTML}{8C2D19}
\definecolor{citecolor}{HTML}{1B4F9C}
\definecolor{urlcolor}{HTML}{0E6B50}
\usepackage[colorlinks=true,linkcolor=linkcolor,citecolor=citecolor,urlcolor=urlcolor]{hyperref}
\makeatletter
\@ifundefined{UrlNoBreaks}{}{\expandafter\def\expandafter\UrlNoBreaks\expandafter{\UrlNoBreaks\do\:}}
\makeatother
\AtBeginEnvironment{table}{\hypersetup{hidelinks}}
\AtBeginEnvironment{thebibliography}{\emergencystretch=2em}

\graphicspath{{figures/}}

\newcommand{\PiP}{\mathrm{PiP}}

\title{Percentile in Percentile: Evaluating Scholars Across Career Stages}
\author{Panos Ipeirotis\thanks{Department of Technology, Operations, and Statistics, Stern School of Business, New York University, New York, NY, USA. ORCID: \url{https://orcid.org/0000-0002-2966-7402}}\textsuperscript{\kern3pt,\kern1pt}\thanks{Corresponding author: \texttt{panos@stern.nyu.edu}}
  \and Sen Tian\thanks{Ciphero, New York, NY, USA.}}
\date{}

\begin{document}
\maketitle

\begin{abstract}
\noindent
Scholars are often evaluated early in their careers, before papers accumulate citations. Standard metrics such as
citation counts and the $h$-index fit these comparisons poorly: they grow with seniority, vary across fields and
inflate over time. Although citation counts take years to build, a paper's citation percentile within its publication
year settles quickly, correlating at 0.91 between ages 3 and 10. We introduce Percentile in Percentile (PiP), a score
rating scholars against their career cohort. At each impact threshold above a cutoff, PiP counts a scholar's papers
reaching that citation percentile, ranks that count against peers who started in the same year, and averages these
standings. Across 11.1 million Semantic Scholar authors, PiP's distribution stays stable across 1960 to 2015 cohorts,
whereas the average $h$-index at career age 10 rose about 40\%. In every setting and across all 15 fields, PiP
outperforms the $h$-index at predicting a scholar's record in the next decade, including future top papers, exceptional
papers and overall citation impact, and reaches its age-10 accuracy in identifying future high performers 1.6 to 1.8
years earlier. Among all eligible authors, PiP also better anticipates later academy memberships, fellowships and
prizes.
\end{abstract}


\section{Introduction}

Hiring, promotion and award committees compare scholars with others at a similar career stage. A committee deciding on
tenure asks whether a candidate's record is strong for someone at that career stage; a funding panel compares
early-career applicants across departments. Bibliometric indicators enter these decisions alongside expert judgment. The
most common are citation counts and the $h$-index \citep{hirsch2005index}, whose values depend on the field, grow with
seniority and drift with the calendar. The average $h$-index ten years into a career is 40\% higher for scholars who
started in 2010 than for those who started in 1960 (Section~\ref{sec:cohorts}). A raw value says little until compared
with peers.

Many of these decisions are made early. Committees for hiring, tenure, fellowships and early-career grants judge
scholars in the first years of their careers, when most of their papers are recent and few have had time to collect many
citations. An indicator is most useful here if, at year five, it already separates the scholars who will do well later
from those who will not. Early detection was one of our motivations.

The $h$-index also reads only part of a record. It is the largest $h$ such that $h$ papers have at least $h$ citations
each, the corner where a scholar's sorted citation curve crosses the diagonal. Papers below the corner do not count,
nor do the citations of papers above it beyond $h$: a scholar with one landmark paper and another with one average
paper both have $h = 1$. A recent paper counts only after collecting $h$ citations, however well it compares with
papers of its own age.

For single papers, percentiles make that comparison. A paper's percentile is the share of papers from its publication
year (and possibly its field) that it out-cites, so a one-year-old paper competes only with other one-year-old papers.
Percentiles have become an established normalization, which the Leiden Manifesto calls the most robust
\citep{bornmann2013use,waltman2016review,hicks2015bibliometrics}. They are less influenced than means by the skewed
tail of the citation distribution \citep{bornmann2013use}, need no assumption about how citations grow with age, and
do not change meaning with the calendar.

Combining paper percentiles into a scholar-level indicator is less settled. Sums such as I3
\citep{leydesdorff2011integrated} grow without bound with the number of papers, so many weak papers can outweigh one
strong paper. Means ignore volume altogether. Counts of papers in the top 10\%
\citep{waltman2012leiden,bornmann2014individual} use one threshold and discard everything below it.

We propose PiP, short for \emph{percentile in percentile}, which places a scholar's papers in percentiles of their
publication year and then places the scholar in percentiles of their career cohort: the peers whose first paper appeared
in the same year. At each impact threshold, we count the scholar's papers at or above that percentile; the share of peers
with fewer such papers is the scholar's standing at that threshold. Across all thresholds, these standings form a curve
(Figure~\ref{fig:curves}). PiP is the average height of this curve above a cutoff. Where the $h$-index reads one corner
of a record, PiP reads the whole curve above the cutoff.

With the cutoff at the 50th percentile, used throughout, only papers in the top half of their year can earn credit, and
recent papers can earn it as soon as they stand out among papers of their own age. A single paper in the top 1\% counts
at every threshold up to the 99th percentile, where most peers have nothing. Lowering the cutoff moves weight toward the
number of papers; raising it, toward their impact. The cutoff is a dial between quantity and impact.

We compute PiP from the Semantic Scholar Academic Graph \citep{kinney2023semantic} for 11.1 million authors at career
ages 1 to 30 (years since their first paper). Authors with at least three articles and an $h$-index of at least three
are \emph{eligible} and enter the comparison. In each evaluation year, we reconstruct peers and percentiles from the
publications and citations up to that year. We make six contributions:
\begin{enumerate}
  \item \textbf{Definition and properties.} At any given impact threshold, a scholar's standing is the share of peers
  with strictly fewer papers above that threshold; PiP averages this standing across all thresholds above the cutoff. It
  is bounded between zero and one, gives nothing to papers strictly below the cutoff, and never decreases when a paper
  is added or its percentile rises (Section~\ref{sec:pip}).
  \item \textbf{Measurement over time.} Paper percentiles settle within three to five years
  (Section~\ref{sec:paperstability}). PiP's distribution changes little across career cohorts from 1960 to 2015
  (Section~\ref{sec:cohorts}). Over a career, a scholar's PiP is more stable than their $h$-index and, over ten-year or
  longer horizons, than their citations (Section~\ref{sec:careerstability}).\looseness=-1
  \item \textbf{Prediction.} PiP predicts future work better than the $h$-index in every setting and field we test,
  whether the $h$-index is raw or a cohort percentile. Top-paper counts, percentile sums and percentile means each
  specialize in one of our three outcomes (the number of future top papers, their citation impact and the chance of an
  exceptional paper), whereas PiP balances all three. Comparing the scholar with peers at every threshold, rather than
  once for the whole record, improves predictions of future citation impact and exceptional papers in every cohort
  (Section~\ref{sec:prediction}).
  \item \textbf{Early detection.} Among eligible authors, PiP identifies future high performers earlier than the
  $h$-index and competing indicators, matching its age-10 accuracy 1.6 to 1.8 years sooner (Section~\ref{sec:early}).
  \item \textbf{Later recognition.} Measured early in a career, PiP anticipates later academy memberships, fellowships
  and prizes better than the $h$-index, with higher cutoffs performing best (Section~\ref{sec:honors}).
  \item \textbf{Robustness.} The results hold without self-citations; PiP rankings from Semantic Scholar agree with
  those from author-curated Google Scholar profiles; and a field-normalized version removes most of the advantage of
  high-citation fields while preserving rankings within each field (Sections~\ref{sec:fields}--\ref{sec:robustness}).
\end{enumerate}

\section{Related work}
\label{sec:related}

\paragraph{The $h$-index and its variants.} The $h$-index \citep{hirsch2005index} combines the number of papers and
their citations into a single number and has been studied, criticized and modified at length. The $g$-index
\citep{egghe2006theory} credits citations above the corner; age-adjusted variants divide by career length or discount
older papers \citep{hirsch2005index,sidiropoulos2007generalized,harzing2014hia}; and \citet{kaur2013universality}
rescale the $h$-index by the field average. Most variants correlate highly with the original
\citep{bornmann2011metaanalysis}. They keep its single-cutoff design, summarizing a scholar's leading core while
ignoring papers ranked beyond it.

\paragraph{Percentiles.} Percentile-based indicators compare each paper with a reference set of papers from the same
year and field \citep{bornmann2013use,leydesdorff2011turning,waltman2013percentile,bornmann2020evaluation}. They are
less sensitive to outliers \citep{bornmann2013use,bornmann2020evaluation} than mean normalization, which compares a
paper's citations with the average of comparable papers \citep{schubert1986relative,Radicchi2008}. The Leiden
Manifesto recommends them \citep{hicks2015bibliometrics}. Ties and the definition of the percentile need care
\citep{waltman2013percentile}. For scoring we use the share of the reference set with strictly fewer citations, so that
uncited papers have percentile zero; for correlations, Hazen's mid-rank \citep{hazen1914storage}
(Section~\ref{sec:properties}).

\paragraph{From papers to scholars.} \citet{bornmann2014individual} propose counting a scholar's papers in the top 10\%
and dividing by career age, and I3 \citep{leydesdorff2011integrated,rousseau2012basic} sums percentiles over a set of
papers. Composite indicators such as that of \citet{ioannidis2019standardized} combine several citation counts, and
\citet{abramo2014productivity} relate field-normalized citations to the cost of labor. For countries and institutions,
double rank analysis expresses each paper's rank within its unit as a function of its rank among all papers in the
world; a power law fitted to this relation estimates the unit's chance of producing a very highly cited paper
\citep{rodriguez2018double,brito2018percentile}. PiP differs from these in normalizing twice: papers by publication
year, and the scholar by career cohort. The cohort comparison covers every impact threshold: the count of a scholar's
papers above each percentile is ranked among career peers, not read against the world distribution of papers.

\paragraph{Prediction of scholarly impact.} \citet{hirsch2007does} found that the $h$-index predicts a scholar's
later $h$-index and citations better than paper or citation counts do. Later work modeled impact as a stable individual
parameter applied to papers of random quality \citep{sinatra2016quantifying}, and \citet{clauset2017datadriven} review
the area. Other work predicted future $h$-index values from career features \citep{acuna2012future} and questioned how
much of that predictability is mechanical, because the future $h$-index contains the current one
\citep{penner2013predictability}. We avoid this overlap by predicting outcomes built solely from papers published after
evaluation.

\paragraph{Citation windows and cumulative advantage.} How long a paper needs before its citations reflect its impact
depends on the field \citep{wang2013window,abramo2011window}. This matters for PiP, which scores recent papers by early
percentiles; Section~\ref{sec:paperstability} measures how fast they settle. Citations also feed on themselves:
well-known scholars receive more attention for the same work \citep{merton1968matthew,petersen2014reputation}. This
cumulative advantage affects any citation-based indicator and any test of its predictive validity
(Section~\ref{sec:discussion}).

\section{Data}
\label{sec:data}

\paragraph{Papers and citations.} We use the 27 October 2026 release of the Semantic Scholar Academic Graph
\citep{kinney2023semantic}, distributed under the ODC-BY license \citep{s2datasets2026}. It contains 226.0 million
papers published from 1950 to 2025. Counting each paper's citations by the year of the citing paper gives its
cumulative citations at each age (years after publication) through 2025. These counts sum to 97.5\% of the
citation counts that Semantic Scholar reports for papers from 2000--2005 (correlation 0.999); the difference consists
of citations from 2026 and from citing papers without a year.

\paragraph{Articles.} Many Semantic Scholar records are not research articles. We keep the 71.5 million papers typed as
journal articles, conference papers, reviews, studies, clinical trials, meta-analyses or case reports, and not as
editorials, news or letters. Only these \emph{articles} enter the reference sets, the paper counts and the $h$-index.
Of the records published in 2000, 63\% are uncited at age 5, but only 30\% of that year's articles. Each article
carries zero or more of Semantic Scholar's 23 fields of study \citep{s2graphapi}. From here on, \emph{papers} means
articles unless stated otherwise.

\paragraph{Authors and careers.} Semantic Scholar assigns author identifiers through automatic disambiguation
\citep{kinney2023semantic}. Disambiguation can split one person across several records or merge different people. We
link authors to their articles (277.8 million links). An author's \emph{start year} is the year of their first
article, and their \emph{career age} at evaluation is the number of years since that start year: an author who started
in 1995 has career age 10 in 2005. We follow every author from career age 1 to 30, or to 2025, whichever comes first.

Every quantity in an evaluation year is computed from the articles published up to that year and their citations up to
that year. This reconstruction is retrospective in one respect: author identities, publication types and fields come from
the 2026 release, not from the records of the time.

\paragraph{Eligibility.} At each evaluation year an author is \emph{eligible} if they have at least three articles and
an $h$-index of at least three, both computed from citations received through that year. The second condition means that
at least three of their articles have at least three citations each, so it implies the first. Because both counts only
grow, an eligible author remains eligible, even after they stop publishing. We compute PiP only for eligible authors
and compare each with their peers, the eligible authors who share their start year. In total, 11.1 million authors are
eligible at some career age, giving 111.6 million author-age records. The $h$-index condition is strict early in a
career and in slow-citing fields: among authors with start years 1975--2000 and at least three articles, it excludes
79\% at career age 3 and 43\% at career age 10, from 22\% in biology to 64\% in economics and 97\% in history and
philosophy. Sections~\ref{sec:early} and~\ref{sec:robustness} repeat the main comparisons under a
\emph{three-article rule}, which requires only the three articles.

\paragraph{Google Scholar profiles.} For validation we use a convenience sample of 16,218 Google Scholar profiles,
mostly of faculty, collected for \url{https://www.pip-score.org} (seeded from US university faculty directories and
expanded with profiles looked up on the site or drawn from Wikidata). Semantic Scholar's author records are built
automatically, although authors can
claim and correct them \citep{kinney2023semantic}. Google Scholar profiles are created and curated by the scholars
themselves \citep{googlescholarprofiles}, so they are a check on automatic disambiguation
(Section~\ref{sec:robustness}). Section~\ref{sec:honors} also uses them in the test of later recognition.

\paragraph{Honors.} For the test of later recognition (Section~\ref{sec:honors}) we use Wikidata \citep{wikidata2026},
a free, collaborative knowledge base whose structured data are released under CC0. Items describing people can record
their honors, with dates, and their Semantic Scholar and Google Scholar identifiers. We queried Wikidata for honors on
5 November 2026; the curated list of 65 senior honors and matching scripts are archived in our replication package
\citep{ipeirotis2026replication}. For the Google Scholar profiles we also record whether the item links to an English
Wikipedia article, using the website's monthly Wikidata copy.

\section{The PiP score}
\label{sec:pip}

A standard way to visualize a scholar's record is the citation-rank graph: papers are sorted from most to least cited,
with raw citations on the vertical axis and paper rank ($k = 1, 2, \ldots$) on the horizontal axis
(illustrated in Figure~\ref{fig:construction}a). The $h$-index summarizes this curve by the largest $h \times h$ square
beneath it ($h = 4$ in Figure~\ref{fig:construction}a). Summarizing a record by that single square has two drawbacks:
geometrically, it ignores both the excess citations of landmark papers above $h$ and all papers ranked beyond $h$;
dynamically, because raw citations take years to accumulate, a recent paper sits near the bottom of the vertical axis
until it collects at least $h$ citations, even if it already leads its publication year (illustrated by article \#2 in
Figure~\ref{fig:construction}a, a hypothetical recent top-10\% paper with three citations, below $h = 4$).

PiP addresses both drawbacks by transforming the two axes in turn: it first normalizes the vertical axis by ranking each
paper against papers from the same year, and then normalizes the horizontal comparison against peers who started in the
same year. We build the score in steps: starting from raw citations against paper rank (Figure~\ref{fig:construction}a),
we replace raw citations with paper percentiles (Figure~\ref{fig:construction}b), normalize the horizontal axis by the
productivity distribution of peers of the same career age (Figure~\ref{fig:construction}c), and show how the limitations
of that normalization motivate comparing the scholar with peers at every impact threshold.

\subsection{Paper percentiles}
Replacing raw citations on the vertical axis with citation percentiles puts papers of different ages on a common
$[0, 1]$ scale: a paper one year old competes only with other papers one year old. At evaluation, an article's
\emph{percentile} is the share of articles from the same publication year with strictly fewer citations. An article at
percentile $0.90$ has more citations than 90\% of articles published in the same year; an uncited article has percentile
$0$. Because the comparison stays within a publication year, paper percentiles need no correction for paper age or for
citation growth over the decades.

Normalizing citations to percentiles also speeds convergence. Whereas raw citation counts take years to build, a paper's
relative standing within its publication year settles quickly: as Section~\ref{sec:paperstability} shows, a paper's
percentile two and three years after publication already correlates at $0.85$ and $0.91$ with its percentile at age ten.
On the percentile scale, a promising recent paper immediately rises near the top of the vertical axis instead of waiting
years to reach $h$ citations (illustrated by article \#2 rising from rank 5 in Figure~\ref{fig:construction}a to rank 2
at $0.925$ in Figure~\ref{fig:construction}b). The main analysis pools fields; Section~\ref{sec:fields} computes
percentiles within fields.

\subsection{From paper rank to productivity-normalized area}
\label{sec:auc}
\begin{figure}[t]
  \centering
  \includegraphics[width=\textwidth]{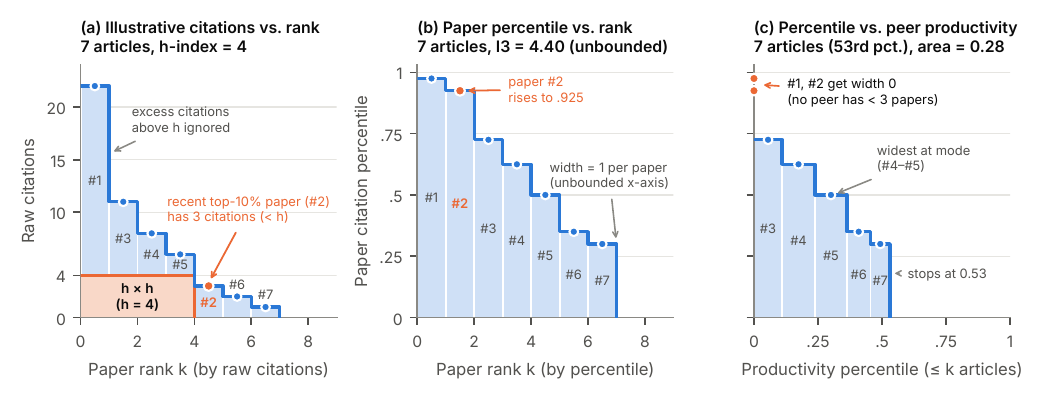}
  \caption{From the $h$-index citation-rank graph to productivity-normalized percentile area, based on the seven-article
  percentile profile of the median eligible author of the 1995 cohort in 2005 (career age 10; articles \#1--\#7 ordered
  by citation percentile). (a) Illustrative raw citations and publication recency against citation rank ($h = 4$): the
  $h \times h$ square ignores excess citations above $h$ (\#1, \#3, \#4, \#5) and papers ranked beyond $h$, including a
  hypothetical recent top-10\% paper (\#2, three citations $< h$) and tail articles (\#6, \#7). (b) Observed within-year
  citation percentiles against paper rank lift article \#2 to $0.925$; each bar has width $1$, so the area equals the
  sum of paper percentiles ($\mathrm{I3} = 4.40$, unbounded). (c) Replacing paper rank $k$ with the share of peers having
  at most $k$ articles bounds the area in $[0, 1]$, ending at the scholar's productivity percentile ($0.53$), but gives
  articles \#1 and \#2 width zero (no eligible peer has fewer than three articles) and the widest bars to middle ranks
  (\#4--\#5).}
  \label{fig:construction}
\end{figure}

With the vertical axis normalized to paper percentiles, a scholar's record can be drawn by sorting their articles from
highest to lowest percentile and plotting paper percentile on the vertical axis against paper rank
($k = 1, 2, 3, \ldots$) on the horizontal axis (Figure~\ref{fig:construction}b). The median scholar of the 1995 cohort at
career age 10, for example, has seven articles: two in the top 10\% of their publication year (ranks 1 and 2), three more
at or above the median (ranks 3 to 5), and two below the median (ranks 6 and 7). Each article forms a bar of width $1$
and height equal to its percentile. The area under this step curve is the sum of the scholar's paper percentiles ($4.40$
in Figure~\ref{fig:construction}b), which is the I3 indicator of \citet{leydesdorff2011integrated}. Because every
additional article adds another bar of width $1$, this area is unbounded: it grows without limit as a scholar publishes
more papers, and a given sum is not comparable across career ages or across cohorts with different publication rates.

A natural way to bound the score and adjust for career stage is to normalize the horizontal axis by the typical
productivity of scholars of the same career age. At the time of evaluation, an author's \emph{peers} are the eligible
authors with the same start year, the author included. Replacing raw paper rank $k$ on the horizontal axis with the share
of peers who have at most $k$ articles in total maps both axes onto $[0, 1]$ (Figure~\ref{fig:construction}c). The curve
now sits inside the unit square and stops horizontally at the author's own productivity percentile ($0.53$ for the
seven-article scholar in Figure~\ref{fig:construction}c). The area under the curve is then a bounded metric in $[0, 1]$
that, for fixed peers, never falls when a paper is added, equaling the curve's horizontal width (the author's
productivity percentile) times its average height (a weighted mean of the author's paper percentiles).

Normalizing the horizontal axis by peers' total paper counts has three drawbacks:
\begin{enumerate}
  \item \emph{The area is capped by total paper count and depends much less on quality.} Because the curve stops on the
  horizontal axis at the scholar's productivity percentile and has height at most $1$, the area can never exceed the
  scholar's percentile on total article count, however cited their papers are. At career age 10, it correlates $0.90$
  with total article count and only $0.13$ with mean paper percentile.
  \item \emph{Bar widths on the horizontal axis are unequal, and early on they increase rather than decrease.} On the
  normalized horizontal axis, the width of the $k$-th article's bar is the share of peers of that career age who have
  \emph{exactly} $k$ articles in total. At high article counts, the right tail of the article-count distribution
  produces diminishing returns, giving each additional article a narrower bar. Early on, however, the width is zero or
  increasing: at any career age past the very beginning, few peers have only one or two articles, while many cluster
  around the typical (modal) article count for that age. Among the eligible authors of a cohort, nobody has fewer than
  three articles, so an author's first and second best articles receive a bar width of zero
  (Figure~\ref{fig:construction}c); among Google Scholar faculty profiles, the median author's top four articles receive
  none. Even past that minimum, as $k$ rises toward the cohort's modal article count, the share of peers with $k$
  articles increases, giving wider bars to the fourth and fifth articles than to the third. Because articles are ordered
  from best to worst along the horizontal axis, this normalization gives zero or narrow bars to a scholar's strongest
  papers and the widest bars to their middle papers. Scored against a simulated cohort of 2,000 eligible peers, an
  author with one article at the 99th percentile gets an area of zero, while an author with 40 articles at the 10th
  percentile gets 0.10.
  \item \emph{Comparing high-impact counts against total counts makes a quality cutoff drift across cohorts.} Read
  horizontally at any impact threshold $\theta$ on the vertical axis, the width of the shaded region in
  Figure~\ref{fig:construction}c is the share of peers whose \emph{total} article count is at most the number of
  articles the author has at or above $\theta$. This area thus compares an author's count of articles above $\theta$
  with peers' total article counts rather than with peers' counts above $\theta$. If we apply a cutoff $\tau$ on the
  vertical axis to ignore low-impact papers, the area compares partial counts above $\tau$ with total counts that grow
  over time, and its median at career age 10 halves between the 1960 and 2010 cohorts.
\end{enumerate}

\subsection{Standing curves and the PiP score}
\label{sec:definition}
These problems disappear once we compare like with like at every impact threshold $\theta \in [0, 1]$. For any threshold
$\theta$, we count how many of the author's articles are at or above percentile $\theta$ of their publication year: at
$\theta = 0$ this is the total number of articles, and at $\theta = 0.90$ it is the number of articles in the top 10\%
of their year. Instead of comparing that count with peers' total articles, we compare it with how many articles peers
have \emph{at or above the same threshold $\theta$}.

\begin{figure}[t]
  \centering
  \includegraphics[width=\textwidth]{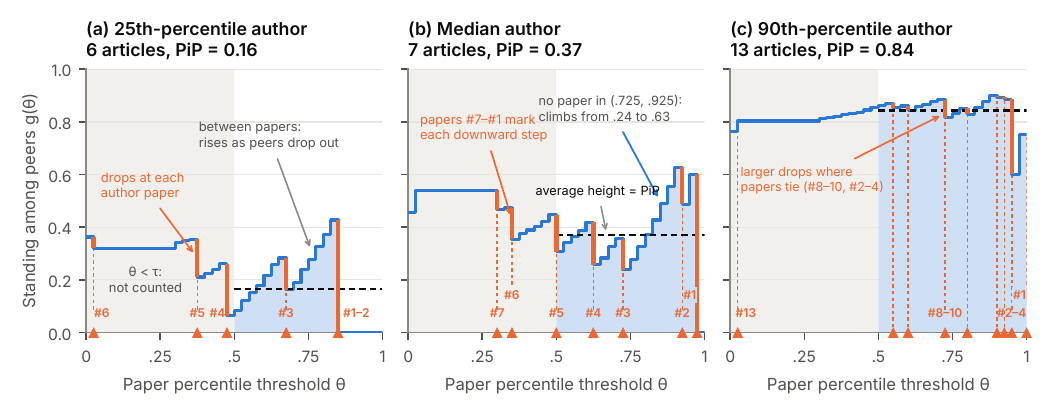}
  \caption{Standing curves of three authors who published their first article in 1995, evaluated in 2005 (career age
  10), chosen as the author whose PiP is closest to the (a) 25th, (b) 50th (median; same author as in
  Figure~\ref{fig:construction}) or (c) 90th percentile of the cohort's scores. Each curve gives, at every threshold
  $\theta$, the share of the 1995 cohort's eligible authors who have fewer articles at or above percentile $\theta$ of
  their publication year. PiP is the shaded area to the right of $\tau = 0.5$, rescaled by $1/(1-\tau)$, or equivalently
  the average height of the curve there (dashed line). Orange triangles and article labels (\#1 = highest citation
  percentile) along the horizontal axis mark each author's articles, and solid orange vertical segments mark the
  downward steps: the curve can fall only where $\theta$ passes one of the author's own articles (with larger drops
  where articles tie), and climbs between articles as peers' articles drop below $\theta$.}
  \label{fig:curves}
\end{figure}

The author's \emph{standing} at threshold $\theta$, written $g(\theta)$, is the share of peers who have strictly fewer
articles at or above $\theta$. For the median author of Figures~\ref{fig:construction} and~\ref{fig:curves}, who has two
articles in the top 10\% ($\theta = 0.90$) and one in the top 5\% ($\theta = 0.95$), 63\% of peers have fewer than two
articles in the top 10\%, so $g(0.90) = 0.63$. An author with no article at or above $\theta$ has $g(\theta) = 0$,
however many peers also have none (Section~\ref{sec:properties}).

Plotted across all thresholds from $\theta = 0$ to $\theta = 1$, the standings trace the author's \emph{standing curve}
$g(\theta)$ (Figure~\ref{fig:curves}). The curve runs from the author's standing on total articles at the left
($\theta = 0$) to their standing on top articles at the right, and need not be monotone: an author can rank higher among
peers on top-10\% articles than on all articles. Visually, the curve rises and falls like a sawtooth. Between two of
the author's articles (marked by orange triangles and labels \#7 to \#1 along the horizontal axis in
Figure~\ref{fig:curves}), raising $\theta$ leaves the number of the author's articles above the threshold unchanged
while some peers' articles drop below it, so the author's standing can only rise. The curve can fall only at an orange
marker where $\theta$ passes one of the author's own articles (with taller drops where multiple articles tie at the same
percentile, as at \#8--10 and \#2--4 for the 90th-percentile author).

The PiP score with cutoff $\tau \in [0, 1)$ is the average height of the standing curve at or above $\tau$:
\begin{equation}
\PiP = \frac{1}{1-\tau} \int_\tau^1 g(\theta)\, d\theta .
\label{eq:pip}
\end{equation}
It is the area under the standing curve to the right of $\tau$, rescaled to $[0, 1]$. The cutoff $\tau$ moves the score
between quantity and impact (Section~\ref{sec:tau}). We use $\tau = 0.5$ throughout unless stated otherwise; with this
cutoff, only articles in the top half of their publication year can earn credit.

The score has a direct head-to-head interpretation. At any given impact threshold $\theta$, $g(\theta)$ is the share of
peers with strictly fewer articles at or above that threshold than the author (a tie is not a win), and PiP averages
that win rate across all thresholds between $\tau$ and $1$. The median author of Figure~\ref{fig:curves} outranks 63\%
of peers when the threshold is at the 90th percentile (the author has two articles there and most peers have fewer),
28\% at the 75th percentile, and none at the 97.5th percentile. Averaged across all thresholds above the cutoff, the
author outranks 37\% of peers ($\PiP = 0.37$), even though their score is near the cohort median. A score of 0.60
therefore means an average win rate of 60\% across thresholds above $\tau$, not being at the 60th percentile of scores.
At career age 10, across cohorts from 1960 to 2015, the lower quartile of PiP lies between 0.16 and 0.20, the median
between 0.37 and 0.41, the upper quartile between 0.64 and 0.68 and the 90th percentile between 0.84 and 0.86
(Section~\ref{sec:cohorts}), so a score of 0.60 lies between the median and the upper quartile of its cohort.

Not counting ties as wins also determines a peer group's mean score. In a head-to-head comparison between any two peers
at a given threshold, either they tie (and neither wins) or one has strictly more articles than the other. Across all
pairs of peers at that threshold, the average win rate is therefore half the share of pairs that do not tie; averaged
over thresholds above $\tau$, the cohort mean of PiP is always below $0.5$ and falls as ties become more common, as at
high cutoffs (Section~\ref{sec:tau}).

To measure what the comparison at every threshold adds, Sections~\ref{sec:prediction} and~\ref{sec:early} also report
\emph{PiP-simple}. It keeps the paper percentiles and the cutoff $\tau$ but compares the scholar with peers only once, by
summing the amount by which each article exceeds $\tau$ (divided by $1-\tau$) and ranking that sum in the cohort.

\subsection{Properties}
\label{sec:properties}
The properties below hold for a fixed peer group containing the author; each follows from the definition
(Appendix~\ref{app:proofs}), and we checked them numerically on simulated cohorts \citep{ipeirotis2026replication}.

\paragraph{Range.} $0 \le \PiP < 1$. Because the peer group includes the author, no one can have more articles than
every peer including themselves, so the standing $g(\theta)$ is strictly below $1$ at every threshold. The score is zero
when every article of the author falls strictly below $\tau$.

\paragraph{Monotonicity.} With the peers and the percentiles of the author's other articles held fixed, adding an
article or raising an existing article's percentile never lowers PiP: at every threshold either change raises or
maintains the author's article count at or above $\theta$, which can only raise or maintain the standing $g(\theta)$.
The gain can be zero: one more article at or above $\theta$ raises $g(\theta)$ only if some peer had the author's
previous count. An article whose percentile stays strictly below $\tau$ affects counts only for $\theta < \tau$ and
leaves the score unchanged. This property holds for paper percentiles rather than raw citations: when a paper gains
citations, its year's percentiles are recomputed, and if a paper ties or overtakes another of the author's own papers
from the same year, it lowers that paper's percentile and can slightly lower the score.

\paragraph{No reward for volume alone.} At $\tau = 0.5$, an author whose articles all fall below the median of their
year scores zero, however many they publish, whereas the productivity-normalized area without a cutoff gives 0.10 to 40
articles at the 10th percentile.

\paragraph{Credit for the best work.} A single highly cited article adds one to the author's count at every threshold up
to its own percentile, including high thresholds where most peers have none. Scored against the simulated cohort of
Section~\ref{sec:auc}, which it does not join, a record of one article at the 97.5th, 99th or 99.9th percentile scores
0.42, 0.45 or 0.47, and one at the 50th percentile scores 0; the productivity-normalized area of
Figure~\ref{fig:construction}c gives all four records zero, and the $h$-index cannot tell them apart (all have $h = 1$).

\paragraph{Ties.} The standing counts peers with strictly fewer articles. Mid-ranks instead count half of the tied
peers. They are a common choice for the percentile of a single paper \citep{bornmann2013use,waltman2013percentile}, but
in $g(\theta)$ they would give credit to authors with nothing above a threshold: at high thresholds most peers have no
article, and an author tied with them at zero would stand at half their share. In the simulated cohort, an author whose
five articles all fall below the median would score 0.24 with mid-ranks instead of 0. The strict rule keeps weak
articles from earning credit through ties; its price is that the mean score of a peer group depends on how common ties
are (Section~\ref{sec:definition}).

\paragraph{Balance of volume and impact.} In the productivity-normalized area of Figure~\ref{fig:construction}c, no
author's score can exceed their percentile on total article count (the horizontal width of their curve). In PiP, by
contrast, comparing counts at each threshold $\theta \ge \tau$ allows an author with fewer, highly cited articles to
outrank peers with more total articles: at $\tau = 0.5$, 30--36\% of eligible authors (depending on career age) achieve
a PiP score above their total-article percentile. At career age 10, whereas the productivity-normalized area correlates
0.90 with total article count and 0.13 with mean paper percentile, PiP ($\tau = 0.5$) balances the two, correlating 0.64
with total article count, 0.60 with top-2.5\% article count and 0.47 with mean paper percentile (Section~\ref{sec:tau}).

\subsection{Relation to other indicators}
\label{sec:relation}
Let $n(\theta)$ be the number of an author's articles at or above percentile $\theta$. Existing indicators and PiP can
all be written in terms of this count curve $n(\theta)$ and the standing curve $g(\theta)$ (Table~\ref{tab:taxonomy}).
The article count reads $n(0)$ and a top-10\% count reads $n(0.90)$, each at a single threshold without a peer
comparison. I3 sums the paper percentiles, which equals $\int_0^1 n(\theta)\,d\theta$: the area under the count curve
without a peer comparison, growing without bound with article count. PiP-simple averages $n(\theta)$ above the cutoff,
$\frac{1}{1-\tau}\int_\tau^1 n(\theta)\,d\theta$, and compares with peers once by rank; PiP replaces the raw count
$n(\theta)$ with the peer standing $g(\theta)$ at every threshold before averaging. The $h$-index reads one corner of
the analogous curve in citation space: the largest integer $h$ where at least $h$ articles have at least $h$ citations.

\begin{table}[t]
  \centering
  \small
  \caption{Indicators as summaries of the count curve $n(\theta)$, the number of the author's articles at or above
  percentile $\theta$, and of the standing curve $g(\theta)$.}
  \label{tab:taxonomy}
  \begin{tabular}{llll}
    \toprule
    Indicator & In terms of the curves & Thresholds used & Comparison with peers \\
    \midrule
    Article count & $n(0)$ & one ($\theta = 0$) & none, or once by rank \\
    Top-10\% count & $n(0.90)$ & one ($\theta = 0.90$) & none, or once by rank \\
    I3 & $\int_0^1 n(\theta)\,d\theta$ & all & none, or once by rank \\
    PiP-simple & $\frac{1}{1-\tau}\int_\tau^1 n(\theta)\,d\theta$ & $\theta \ge \tau$ & once, by rank \\
    PiP & $\frac{1}{1-\tau}\int_\tau^1 g(\theta)\,d\theta$ & $\theta \ge \tau$ & at every threshold \\
    \bottomrule
  \end{tabular}
\end{table}

\paragraph{Numerical evaluation.} To compute PiP in our empirical analyses, we divide $[0, 1)$ into 40 percentile bins
of width $0.025$. At the default cutoff $\tau = 0.5$, the reported score averages the author's standing $g(\theta)$ over
the 20 thresholds $0.50, 0.525, \ldots, 0.975$ (which equals the exact integral in Eq.~\ref{eq:pip} when each paper
percentile is rounded up to the top of its bin; Appendix~\ref{app:proofs}). Against the unrounded integral at career
age 10, the within-cohort rank correlation is 0.9994 and top-decile overlap is approximately 96\%; predictive statistics
differ by at most 0.004 in the tested comparisons. The replication package documents the computation, convergence
checks, and finer-grid calculations used for the illustrative examples.

\section{Results}
\label{sec:results}

We first examine PiP as a measurement: how soon a paper's percentile settles, what the cutoff changes, whether a score
is equally rare in every cohort and how much it moves over a career. We then test whether PiP anticipates future work
and later honors, and check the findings by field and under other data choices.

\subsection{Paper percentiles settle within a few years}
\label{sec:paperstability}

PiP uses percentiles of recent papers, measured after only a few years of citations, so we check how well an early
percentile anticipates the later one. For every article published from 1980 to 1995, we compute its percentile within
its publication year at each age, correlate the percentiles at two ages within each year and average over publication
years. These percentiles are not the score's strict shares but Hazen mid-ranks \citep{hazen1914storage}, so the
correlation is the Spearman correlation of the citation counts.

\begin{figure}[t]
  \centering
  \includegraphics[width=\textwidth]{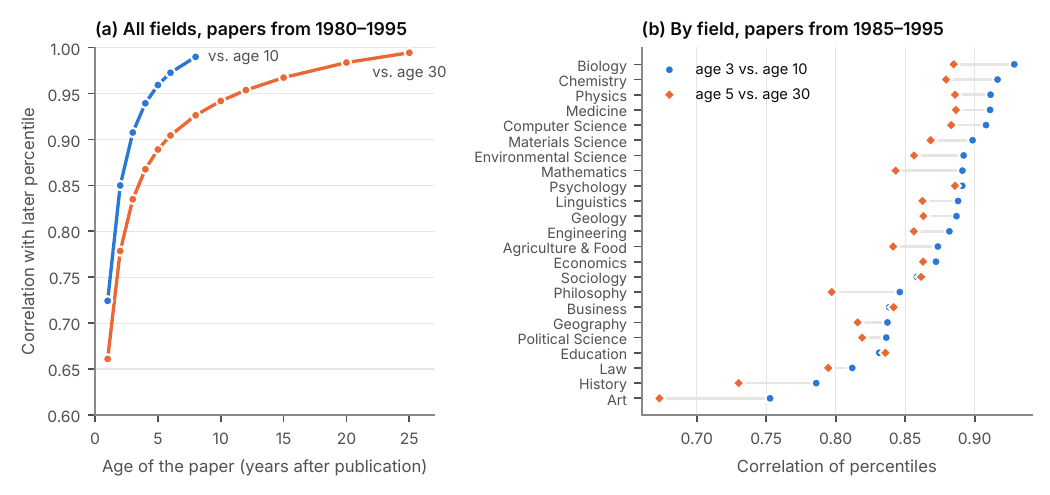}
  \caption{How well a paper's early citation percentile anticipates its later percentile. (a) Correlation between
  a paper's percentile at an early age and at age 10 or 30, for 7.2 million articles published in 1980--1995, averaged over
  publication years. (b) The same correlation by field, for articles published in 1985--1995, with percentiles
  computed within the field and publication year; fields with more than 20,000 articles.}
  \label{fig:paperstability}
\end{figure}

Percentiles settle quickly (Figure~\ref{fig:paperstability}a, Table~\ref{tab:paperstability}). The correlation with age
10 is 0.72 after one year, 0.91 after three and 0.96 after five; with age 30 it is 0.84 after three and 0.89 after five.
The ordinary least-squares slope of the later on the earlier percentile is close to one (0.94 for ages 3 and 10),
so there is little regression to the mean: early high-percentile papers stay near the top rather than drifting
toward the middle. Individual papers still move: between ages 3 and 10 a paper's percentile changes by 8.6
points (0.086) on average.

The agreement between ages 3 and 10 holds in every publication year from 1960 to 1995: the correlation lies between
0.875 and 0.912. Early percentiles have also become more informative: the correlation between ages 1 and 10 was 0.72
for publication years 1980--1995 and 0.83 for 2015.

\begin{table}[t]
  \centering
  \small
  \caption{Agreement between a paper's citation percentile at an early and a later age (articles published 1980--1995,
  7.2 million articles). Correlation of Hazen percentiles within publication year, the slope of the later on the earlier
  percentile, and the mean absolute change.}
  \label{tab:paperstability}
  \begin{tabular}{lccc}
    \toprule
    Ages & Correlation & Slope & Mean absolute change \\
    \midrule
    1 vs.\ 10 & 0.724 & 0.846 & 0.161 \\
    2 vs.\ 10 & 0.850 & 0.911 & 0.115 \\
    3 vs.\ 10 & 0.908 & 0.944 & 0.086 \\
    5 vs.\ 10 & 0.959 & 0.975 & 0.051 \\
    3 vs.\ 30 & 0.835 & 0.882 & 0.125 \\
    5 vs.\ 30 & 0.889 & 0.918 & 0.100 \\
    10 vs.\ 30 & 0.942 & 0.956 & 0.068 \\
    20 vs.\ 30 & 0.984 & 0.989 & 0.030 \\
    \bottomrule
  \end{tabular}
\end{table}

The speed differs by field (Figure~\ref{fig:paperstability}b), in line with \citet{wang2013window} and
\citet{abramo2011window}. Three years give a correlation of about 0.9 with age 10 in biology (0.93), chemistry (0.92),
physics, medicine and computer science (0.91), but 0.79 in history and 0.75 in art. The humanities and law need about
five years to reach the age-10 agreement that the natural and life sciences reach in three (at age 5, correlation with
age 10 reaches 0.92 in philosophy, 0.91 in law, 0.88 in history and 0.86 in art). PiP uses each paper's percentile at
its current age, so in slow fields part of a scholar's score rests on unsettled percentiles;
Section~\ref{sec:discussion} discusses what this means for scholars in those fields.

\subsection{The cutoff \texorpdfstring{$\tau$}{tau} is a dial between quantity and impact}
\label{sec:tau}

\begin{figure}[t]
  \centering
  \includegraphics[width=\textwidth]{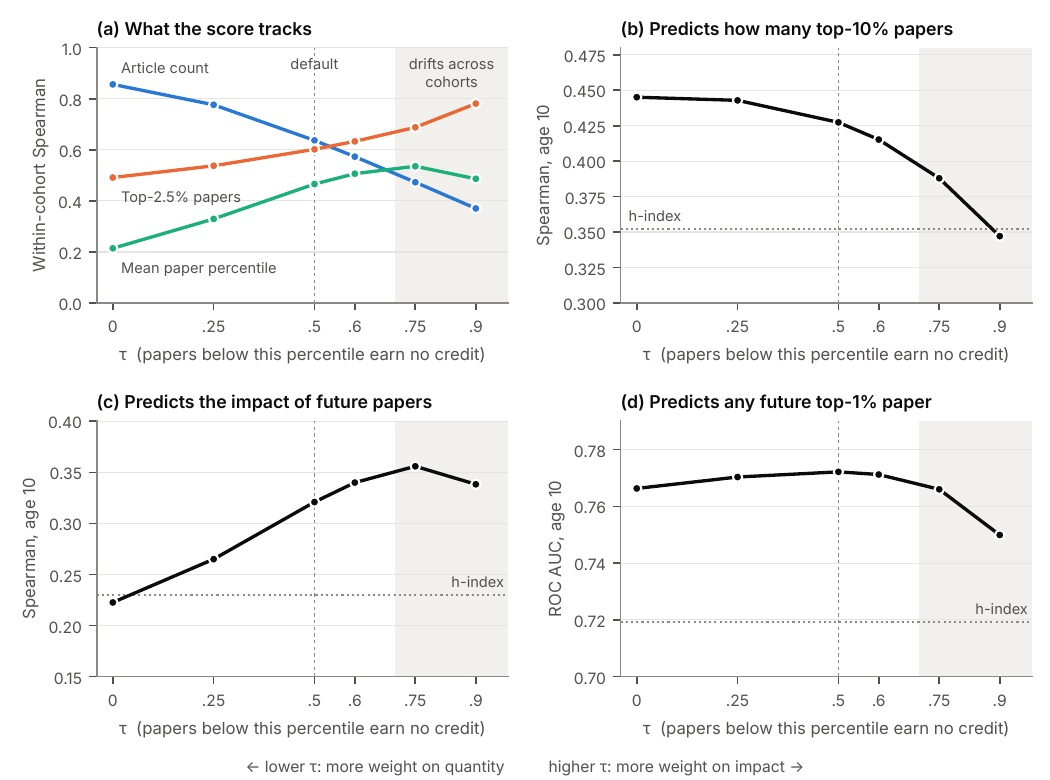}
  \caption{The cutoff $\tau$ moves PiP from quantity toward impact. Eligible authors at career age 10, start years
  1975--2010. (a) Within-cohort Spearman correlation of PiP with the author's number of articles, number of articles in
  the top 2.5\% of their year, and mean article percentile. (b--d) Prediction of the author's articles published in the
  next ten years, for start years 1975--2000 (Section~\ref{sec:prediction}): (b) Spearman correlation with the number of
  future top-10\% articles, (c) Spearman correlation with their mean percentile, (d) ROC AUC for publishing at least one
  top-1\% article. The horizontal line shows the $h$-index. In the shaded range the score's distribution drifts across
  cohorts.}
  \label{fig:tau}
\end{figure}

The cutoff decides which papers earn credit. To see how it moves the score between quantity and impact, we computed
PiP at cutoffs $\tau = 0, 0.25, 0.5, 0.6, 0.75$ and $0.9$ for 3.43 million eligible authors at career age 10 (start
years 1975--2010). Within each cohort, we correlated each version with the author's article count, mean article
percentile and top-2.5\% article count (Figure~\ref{fig:tau}a, Table~\ref{tab:tau}).

\begin{table}[t]
  \centering
  \footnotesize
  \setlength{\tabcolsep}{4.5pt}
  \begin{threeparttable}
  \caption{PiP at different cutoffs $\tau$, career age 10. Columns 2--4: within-cohort Spearman correlation with
  characteristics of the record (3.43 million authors, start years 1975--2010). Columns 5--7: prediction of the next
  ten years (1.45 million authors, start years 1975--2000). Last column: median score of the 1960 and 2010 cohorts.}
  \label{tab:tau}
  \begin{tabular}{lccccccc}
    \toprule
    & \multicolumn{3}{c}{Correlation with} & \multicolumn{3}{c}{Prediction} & \\
    \cmidrule(lr){2-4}\cmidrule(lr){5-7}
    $\tau$ & Articles & Mean pct. & Top 2.5\% & Top-10\% count & Top-1\% AUC & Mean pct. & Median \\
    \midrule
    0 & 0.856 & 0.214 & 0.491 & \textbf{0.445} & 0.766 & 0.223 & 0.44 $\to$ 0.40 \\
    0.25 & 0.776 & 0.329 & 0.537 & 0.443 & 0.770 & 0.265 & 0.43 $\to$ 0.39 \\
    \textbf{0.5} & 0.637 & 0.466 & 0.602 & 0.427 & \textbf{0.772} & 0.321 & 0.41 $\to$ 0.38 \\
    0.6 & 0.573 & 0.506 & 0.633 & 0.415 & 0.771 & 0.340 & 0.41 $\to$ 0.37 \\
    0.75 & 0.473 & \textbf{0.536} & 0.688 & 0.388 & 0.766 & \textbf{0.356} & 0.38 $\to$ 0.33 \\
    0.9 & 0.370 & 0.486 & \textbf{0.781} & 0.347 & 0.750 & 0.338 & 0.37 $\to$ 0.16 \\
    \midrule
    $h$-index & -- & -- & -- & 0.352 & 0.719 & 0.230 & -- \\
    \bottomrule
  \end{tabular}
  \end{threeparttable}
\end{table}

As $\tau$ rises, the correlation with the number of articles falls from 0.86 to 0.37, that with the number of
top-2.5\% articles rises from 0.49 to 0.78 and that with the mean percentile peaks at $\tau = 0.75$. The same dial
trades quantity for impact when PiP predicts the author's future articles (Section~\ref{sec:prediction}): the
\emph{number} of future top-10\% papers is best predicted at low $\tau$, and the \emph{citation impact} of future
articles (their mean percentile) at $\tau = 0.75$; the chance of a future top-1\% paper is predicted about equally well
from $\tau = 0.25$ to $0.6$.

Above $\tau = 0.6$ the score's distribution drifts across cohorts through ties at zero (Section~\ref{sec:definition}):
from 1960 to 2010, the share of eligible authors scoring zero rises from under 0.1\% to 15\% at $\tau = 0.75$ (halving
the lower quartile from 0.19 to 0.08) and from 6\% to 42\% at $\tau = 0.9$, pulling the median from 0.37 to 0.16
(Table~\ref{tab:tau}).

We use $\tau = 0.5$ as the default: it has a one-line explanation (only papers in the top half of their year count),
balances the correlations with quantity (0.64) and with top papers (0.60), best predicts a future top-1\% paper and
keeps the score's distribution stable across cohorts. Users who favor volume can lower it, and those who favor impact
can raise it up to about 0.6, chosen and stated in advance.

\paragraph{Choosing the cutoff on earlier cohorts.} Because $\tau = 0.5$ was selected from results that pooled all
cohorts, we checked whether choosing the cutoff on earlier cohorts alone gives the same answer. We split the
same-career-age prediction settings into a choice sample (start years 1975--1990) and a test sample (start years 1991
and later). Stated as a rule, our criterion picks the cutoff with the highest AUC for a future top-1\% paper among
cutoffs whose distribution is stable across cohorts ($\tau \le 0.6$). Applied to the choice sample alone, that rule
selects $\tau = 0.6$ at career age 10, with $\tau = 0.5$ within 0.001. The choice cohorts differ slightly from the test
cohorts: the knee of their quantity--impact frontier (Figure~\ref{fig:frontier}) lies between 0.6 and 0.75 rather than
at 0.5, and at career age 5 the top-10\% count ties $\tau = 0.5$ on the top-1\% outcome and beats it on future citation
impact, whereas $\tau = 0.6$ keeps its lead.

In the test cohorts both cutoffs keep every comparison reported in Sections~\ref{sec:prediction}--\ref{sec:early} at
career ages 5, 10 and 15: they beat the $h$-index on all three outcomes, no other indicator beats either of them on more
than one outcome, and both have the highest AUC for a future top-1\% paper. Across $\tau$ from 0.25 to 0.6 that AUC
varies by less than 0.01 in the test cohorts, so the exact choice in this range matters little for that outcome.

\subsection{The distribution of PiP is stable across cohorts}
\label{sec:cohorts}

\begin{figure}[t]
  \centering
  \includegraphics[width=\textwidth]{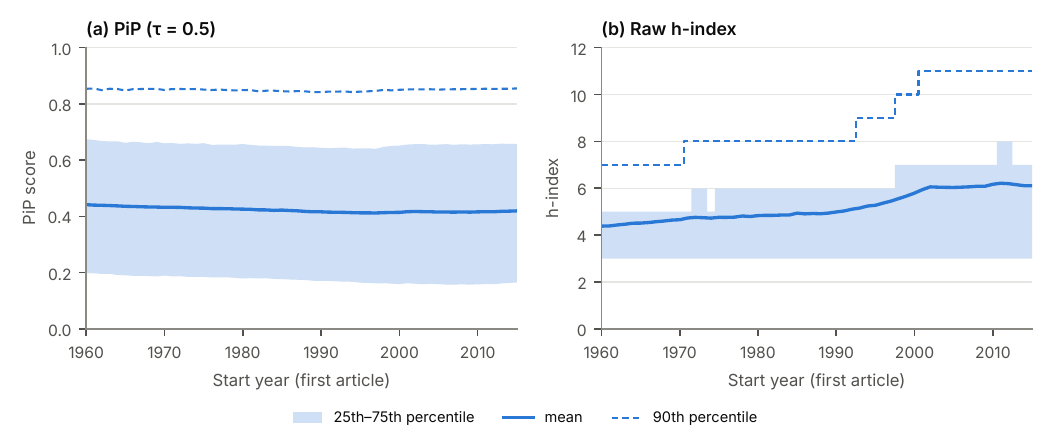}
  \caption{Distribution of (a) PiP ($\tau = 0.5$) and (b) the raw $h$-index of eligible authors at career age 10, by
  start year. The PiP distribution changes little over 56 cohorts; the mean $h$-index grows by about 40\%.}
  \label{fig:cohorts}
\end{figure}

Committees compare people who started in different years, and indicators are often read against fixed rules of thumb.
A fixed rule treats every start year alike only if the indicator's distribution does not change with the start year.
Figure~\ref{fig:cohorts} shows the distributions of PiP and the raw $h$-index at career age 10 for every start year
from 1960 to 2015. The cohorts grow nearly a hundredfold, from 4,031 eligible authors in 1960 to 392,360 in 2015. The
PiP distribution changes little: the mean goes from 0.44 to 0.42, the lower quartile from 0.20 to 0.17, the upper
quartile from 0.67 to 0.66 and the 90th percentile stays at 0.85. Over the same cohorts the mean $h$-index at age 10 rises
by 40\%, from 4.4 in the 1960 cohort to 6.2 in 2010 (6.1 in 2015), and its 90th percentile from 7 to 11, following the
growth of publication and citation rates.

Part of this stability is built in: because a peer group's mean score equals half the share of peer pairs that do not
tie above $\tau$ (Section~\ref{sec:definition}), the mean can move only as far as tie rates change. The rest of
the distribution can still drift when ties at zero become common, as happens above $\tau = 0.6$
(Section~\ref{sec:tau}).

Because the distribution is stable across cohorts, a PiP of 0.6 at career age 10 was about equally rare in the 1970 and
2010 cohorts, so raw scores at a given career age can be compared directly across start years without converting them
into cohort percentiles. Section~\ref{sec:prediction} tests whether raw scores also predict future work across pooled
cohorts and seniorities.

\subsection{Stability over a career}
\label{sec:careerstability}

An indicator that reshuffles scholars every few years is of little use for decisions with long consequences. To measure
this reshuffling, we correlate each indicator's within-cohort percentile at two career ages, for authors who started in
1975--1995 (Table~\ref{tab:careerstability}). Over horizons starting early in a career, PiP is much more stable than
the $h$-index: between ages 5 and 30 its correlation is 0.56, against 0.37 for the $h$-index and 0.45 for citations.
The $h$-index is particularly unstable early on, partly because many eligible authors share the values 3 and 4.
Citations are slightly more stable than PiP between ages 5 and 10 and less stable over longer horizons. Article
count is more stable than PiP at every horizon, because past papers never change their count whereas new citations
continue to re-rank existing work.

\begin{table}[t]
  \centering
  \small
  \caption{Stability of scholar indicators over a career. Correlation between an indicator's within-cohort percentile
  at two career ages (equal to the Spearman correlation of the raw values within cohort); eligible authors with start
  years 1975--1995. The 5-vs-30 comparison covers 434,000 authors.}
  \label{tab:careerstability}
  \begin{tabular}{lccccc}
    \toprule
    Indicator & 5 vs.\ 10 & 5 vs.\ 15 & 5 vs.\ 30 & 10 vs.\ 30 & 20 vs.\ 30 \\
    \midrule
    Articles & 0.847 & 0.754 & 0.642 & 0.800 & 0.946 \\
    \textbf{PiP} ($\tau = 0.5$) & 0.781 & 0.687 & 0.563 & 0.726 & 0.911 \\
    Citations & 0.798 & 0.659 & 0.451 & 0.620 & 0.886 \\
    $h$-index & 0.604 & 0.496 & 0.370 & 0.599 & 0.877 \\
    \bottomrule
  \end{tabular}
\end{table}
\subsection{Predicting future work}
\label{sec:prediction}

The hardest test of an indicator for career decisions is whether it anticipates the scholar's next work. For each
eligible author, we score each article they publish in the ten years after the evaluation year by its percentile within
its publication year at its own age 5 (at age 5 a paper's percentile already correlates at 0.96 with its age-10
percentile; Section~\ref{sec:paperstability}). Indicators use only the author's articles and citations up to the
evaluation year, so no article counts toward both predictor and outcome. We use three outcomes, each with its own
accuracy measure: the number of future articles in the top 10\% (Spearman correlation); whether at least one future
article is in the top 1\% (ROC AUC: here and below, the probability that a random author with the outcome outranks a
random author without it, counting ties as one half, so 0.5 is chance); and the citation impact of the future
articles, defined as their mean percentile, among authors with at least one (Spearman correlation). We compare PiP with
the $h$-index, total citations, the article count and several indicators built from the same paper percentiles
(defined below).

We evaluate predictors in two ways that mirror how committees compare scholars, giving five settings in total.
\emph{Same career age} (as for a tenure committee comparing candidates of similar seniority) pools all eligible authors
at career age 5, 10 or 15 across start years from 1975 up to 2005, 2000 or 1995 (that is, $2010$ minus the career age,
so every future article has five years of citations by 2025). \emph{Same year} (as for a panel comparing candidates of
different seniority) pools all eligible authors in 2000 or 2005 at career ages 2 to 30. Because Spearman correlation
and ROC AUC depend only on rank order, a raw indicator and its cohort percentile rank authors identically within a
single cohort; across cohorts they differ, so in the same-year settings we compare each indicator in both raw and
cohort-percentile form.

\begin{figure}[t]
  \centering
  \includegraphics[width=\textwidth]{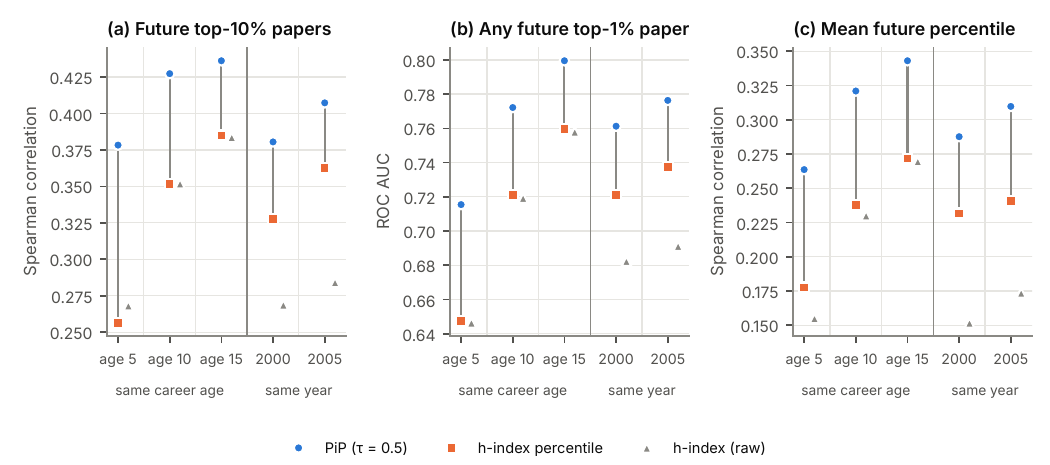}
  \caption{Prediction of the next ten years of a scholar's work from indicators at the evaluation year. Settings:
  eligible authors at career age 5, 10 or 15 (left of the vertical line), or all eligible authors in 2000 or 2005 at
  career ages 2--30 (right). Outcomes: (a) Spearman correlation with the number of future articles in the top 10\% of
  their year; (b) ROC AUC for at least one future top-1\% article; (c) Spearman correlation with the mean percentile
  of future articles. Each future article is scored at age 5. Gray segments join raw PiP to the cohort percentile of
  the $h$-index.}
  \label{fig:prediction}
\end{figure}

\begin{table}[t]
  \centering
  \small
  \begin{threeparttable}
  \caption{Prediction of the next ten years of work. Career age 10 (1.45 million authors, start years 1975--2000; raw
  values, pooled over start years) and all eligible authors in 2005 (1.85 million authors, career ages 2--30;
  within-cohort percentiles, except where noted). Spearman correlation with the number of future top-10\% articles, ROC
  AUC for at least one future top-1\% article, and Spearman correlation with the mean percentile of future articles. The
  best value in each column is in bold. PiP-simple (Section~\ref{sec:definition}) compares the scholar with peers only
  once.}
  \label{tab:prediction}
  \begin{tabular}{lcccccc}
    \toprule
    & \multicolumn{3}{c}{Career age 10, raw} & \multicolumn{3}{c}{Year 2005, cohort pct.} \\
    \cmidrule(lr){2-4}\cmidrule(lr){5-7}
    Indicator & Top-10\% & Top-1\% & Mean pct. & Top-10\% & Top-1\% & Mean pct. \\
    \midrule
    \textbf{PiP} ($\tau = 0.5$) & 0.427 & \textbf{0.772} & 0.321 & 0.416 & \textbf{0.781} & 0.318 \\
    \quad PiP-simple ($\tau = 0.5$) & 0.430 & 0.765 & 0.296 & 0.422 & 0.778 & 0.296 \\
    \addlinespace
    $h$-index & 0.352 & 0.719 & 0.230 & 0.363 & 0.737 & 0.241 \\
    \quad raw, in 2005 & & & & 0.284 & 0.691 & 0.174 \\
    Citations & 0.290 & 0.714 & 0.298 & 0.312 & 0.733 & 0.303 \\
    Articles & 0.437 & 0.740 & 0.000 & \textbf{0.439} & 0.757 & 0.019 \\
    \addlinespace
    Top-10\% count & 0.360 & 0.748 & 0.338 & 0.354 & 0.762 & 0.324 \\
    \quad per year of career, raw & & & & 0.379 & 0.771 & 0.345 \\
    Top-1\% count & 0.250 & 0.641 & 0.228 & 0.262 & 0.698 & 0.225 \\
    I3 (sum of percentiles) & \textbf{0.438} & 0.755 & 0.182 & 0.434 & 0.769 & 0.189 \\
    Mean percentile & $-0.043$ & 0.523 & \textbf{0.394} & $-0.042$ & 0.526 & \textbf{0.404} \\
    \bottomrule
  \end{tabular}
  \end{threeparttable}
\end{table}

\begin{figure}[t]
  \centering
  \includegraphics[width=\textwidth]{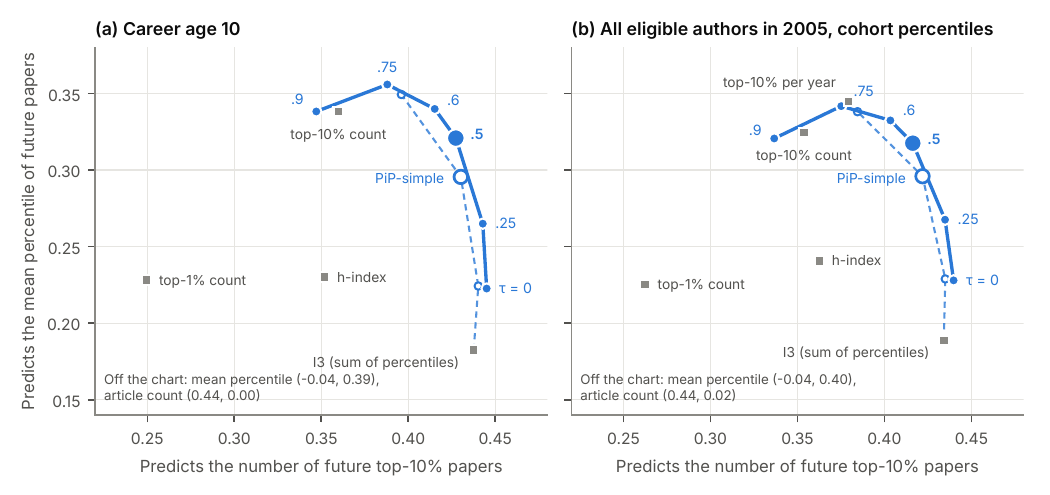}
  \caption{Indicators placed by how well they predict the number of a scholar's future top-10\% papers (horizontal) and
  the mean percentile of their future papers (vertical), Spearman correlations. The blue curve is PiP at cutoffs from
  $\tau = 0$ to $0.9$; the large dot is $\tau = 0.5$. Gray squares are the $h$-index, I3, the top-10\% and top-1\%
  counts and, in panel (b), the top-10\% count per year of career; the mean percentile and the article count lie off
  the chart, and their coordinates are printed in each panel. (a) Eligible authors at career age 10, raw values. (b) All
  eligible authors in 2005 at career ages 2--30, within-cohort percentiles (the top-10\% count per year of career is
  raw). The dashed curve is PiP-simple from $\tau = 0$ (I3) to $0.75$, with the large open circle at $\tau = 0.5$. For
  the third outcome, a future top-1\% paper, PiP at every cutoff from 0 to 0.75 has a higher AUC than every gray
  indicator in both settings.}
  \label{fig:frontier}
\end{figure}

PiP beats the $h$-index on all three outcomes, in all five settings (Figure~\ref{fig:prediction},
Table~\ref{tab:prediction}). At career age 10, its correlation with the number of future top-10\% papers is 0.43 against
0.35, its AUC for a future top-1\% paper 0.77 against 0.72, and its correlation with future citation impact 0.32
against 0.23. The gap is larger when seniority differs: in 2005 the raw $h$-index reaches 0.28, 0.69 and 0.17, against
0.41, 0.78 and 0.31 for raw PiP. Converting the $h$-index into a cohort percentile helps (0.36, 0.74, 0.24) but does
not close the gap. PiP gains much less from conversion: in 2005 its cohort percentile reaches 0.42, 0.78 and 0.32,
because each raw score is already computed within the author's start-year cohort and its scale shifts little across
career ages. Table~\ref{tab:prediction} and Figure~\ref{fig:frontier}b report cohort percentiles for every indicator in
the same-year settings; in the same-career-age settings the two forms nearly agree, and Table~\ref{tab:predictionall}
in the appendix gives raw PiP and both $h$-index forms in all five settings and under the three-article rule.

\paragraph{Productivity and citations.} Because productive authors keep producing, the article count predicts the
number of future top-10\% papers as well as PiP or slightly better (0.44 against 0.43 at age 10), but not future
citation impact (correlation 0.00). Total citations predict future citation impact about as well as PiP (0.30 against
0.32) but are weaker on the number of future top papers (0.29).

\paragraph{Percentile-based indicators.} The closest alternatives to PiP combine the same paper percentiles differently:
the number of papers in the top 10\% \citep{waltman2012leiden}, also divided by career age
\citep{bornmann2014individual}; the number in the top 1\%; I3, the sum of the percentiles
\citep{leydesdorff2011integrated}; and their mean. Apart from the top-1\% count, which trails PiP on all three outcomes,
each is best or nearly best on one outcome and weak on another (Table~\ref{tab:prediction}). The mean percentile
predicts future citation impact best (0.39 at age 10, against 0.32 for PiP), but says nothing about how many top papers
will follow ($-0.04$) or whether one will be in the top 1\% (AUC 0.52). I3 predicts the number of future top papers as
well as the article count does (0.44) but not their citation impact (0.18). The top-10\% count predicts future citation
impact slightly better than PiP (0.34 against 0.32) and the number of future top papers clearly worse (0.36 against
0.43). In all five settings, each of these indicators beats PiP on at most one of the three outcomes, and none on the
chance of a future top-1\% paper. PiP best predicts a future top-1\% paper in every setting; the top-10\% count, I3
and the mean percentile each serve one outcome, whereas PiP balances all three.

Plotting prediction of future top-paper count (horizontal) against prediction of future citation impact (vertical)
reveals a trade-off frontier (Figure~\ref{fig:frontier}). As $\tau$ moves from $0$ to $0.9$, PiP traces a curve that
lies on or outside the other ways of combining paper percentiles, apart from the two single-purpose extremes (the mean
percentile and the article count). With $\tau = 0$, PiP predicts at least as well as I3 on all three outcomes in every
setting (with $\tau = 0.25$, in all but one); with $\tau = 0.75$ it does at least as well as the top-10\% count on all
three outcomes in every setting; and from $\tau = 0.25$ to $0.75$ it beats the $h$-index on all three. Only three
indicators touch or cross the curve: the mean percentile in the direction of pure citation impact (where it gives up
all information about quantity), the article count just beyond the pure-quantity end in two settings (at age 15 and in
2000), and the top-10\% count per year of career, which differs from the plain count only when seniorities are mixed
and sits on the curve next to $\tau = 0.75$. From $\tau = 0$ to $0.75$, choosing $\tau$ picks a point on this frontier
(at $\tau = 0.9$, ties at zero pull both predictions down), and $\tau = 0.5$ sits near its knee: raising $\tau$ from
$0.25$ to $0.5$ gains 0.056 in future citation impact at a cost of 0.016 in quantity, whereas raising $\tau$ past $0.5$
costs much more in quantity for each gain in impact.

\paragraph{Replication across cohorts.} With more than a million authors per setting, sampling error is small: the
standard error of a Spearman correlation of 0.4 would be below 0.001 for independent observations. Authors share
papers, so the observations are not independent and this standard error is optimistic. The more relevant question is
whether the differences hold in different populations. We repeated the age-10 comparison separately for each start
year from 1975 to 2000, 26 cohorts of 17,867 to 122,114 authors. PiP beats the $h$-index on all three outcomes in all
26 cohorts, by 0.06--0.10 on the number of future top papers. It beats the top-10\% count on the number of future top
papers (by 0.05--0.08) and on the top-1\% outcome in all 26 cohorts, and falls behind it on future citation impact in
20 (by up to 0.07). Against I3 and the article count, PiP wins on the top-1\% outcome and on future citation impact in
every cohort, and on the number of future top papers in 12 and 11 of the 26, with differences between $-0.03$ and
$+0.04$.

\paragraph{Comparing with peers at every threshold.} PiP-simple (Section~\ref{sec:definition}) uses the same percentiles
and cutoff as PiP but compares the scholar with peers only once, through the rank of its score in the cohort. The pair
therefore shows what the comparison at every threshold contributes. In each of the 26 cohorts at career age 10, that
comparison improves the AUC for a future top-1\% paper (by 0.004 to 0.011) and the prediction of future citation impact
(by 0.023 to 0.036); the same holds in every cohort at ages 5 and 15, while on future top-10\% papers the two stay
within 0.01. Across $\tau = 0$ to $0.75$, PiP-simple stays inside PiP's frontier (Figure~\ref{fig:frontier}), trailing
by 0.02--0.04 on future impact at $\tau \le 0.25$ and by 0.004--0.011 on top-1\% AUC up to $\tau = 0.5$ before
converging near $\tau = 0.75$. Mixing cohort percentiles of article count and mean percentile also falls short (at age
10: $0.33, 0.72, 0.32$ at $50/50$; $0.44, 0.75, 0.14$ at $75/25$; $0.10, 0.61, 0.40$ at $25/75$).

\subsection{Early detection}
\label{sec:early}

Early detection asks how soon each indicator picks out the scholars who will do well later. We therefore fix the
population and the outcome and vary only the measurement age. The population is every author eligible at career age 3
(306,854 authors, start years 1975--2000); because eligibility persists, each has every indicator at every later age.
The outcome is their work at career ages 11--20, scored as before (each article at its own age 5). A \emph{future star},
or future high performer, is an author in the top 10\% of their start-year cohort, within the population, by their number
of top-10\% articles at ages 11--20 (ties broken by mean percentile).

We measure each indicator at career ages 3, 4, \ldots, 10 as a cohort percentile and ask how well it finds future
stars. The first measure is precision: the share of future stars among the top 10\% of each cohort by the indicator
(chance is 0.10). When authors tie at the 90th-percentile boundary (as often happens with the integer-valued $h$-index),
they split the remaining top-10\% slots equally, which gives the expected precision under random tie-breaking. The
second is the AUC for future stars, and the third the AUC for at least one top-1\% article at ages 11--20.

\begin{figure}[t]
  \centering
  \includegraphics[width=\textwidth]{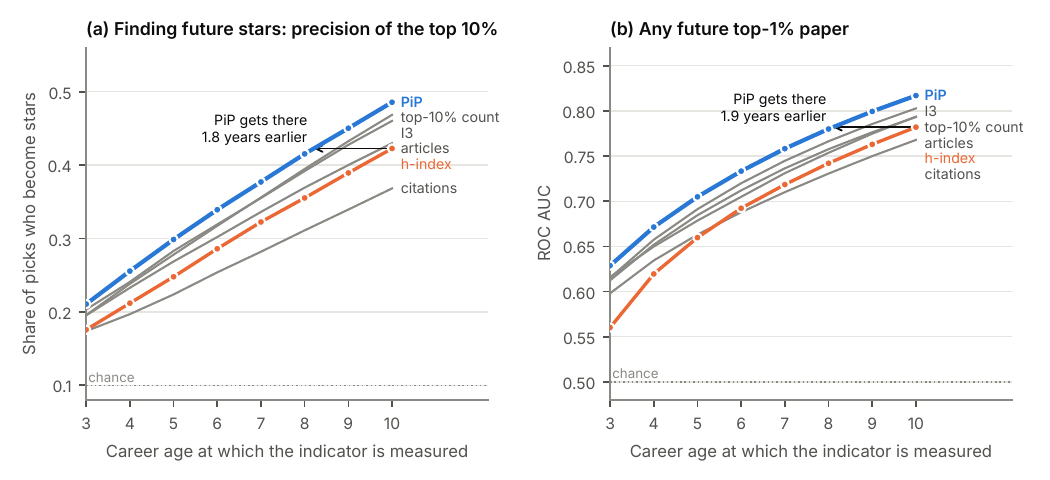}
  \caption{Early detection. Authors eligible at career age 3 (306,854; start years 1975--2000), with each indicator
  measured at career ages 3 to 10 and the outcome fixed: their articles published at career ages 11--20. (a) Of the
  authors in the top 10\% of their cohort by the indicator (ties at the boundary shared at random), the share who
  become future stars (the top 10\% of the cohort by the number of top-10\% articles at ages 11--20). (b) ROC AUC for
  at least one top-1\% article at ages 11--20. The arrow marks the career age where PiP matches the age-10 accuracy of
  the $h$-index.}
  \label{fig:early}
\end{figure}

\begin{table}[t]
  \centering
  \small
  \caption{How much earlier PiP reaches each indicator's accuracy, in years. For each target age from 5 to 10, we find
  the (interpolated) age at which PiP first matches the indicator's accuracy at that target age; the table gives the
  range of lead times across target ages 5 to 10. Authors eligible at career age 3, outcome at career ages 11--20.}
  \label{tab:early}
  \begin{tabular}{lccc}
    \toprule
    Compared with & Precision for future stars & AUC for future stars & AUC for a future top-1\% paper \\
    \midrule
    Citations & 1.7--3.2 & 1.4--2.8 & 1.2--2.6 \\
    $h$-index & 1.2--1.8 & 1.1--1.6 & 1.3--1.9 \\
    Articles & 0.7--1.6 & 0.3--0.9 & 0.6--1.3 \\
    Top-10\% count & 0.5--0.6 & 1.0--1.3 & 0.8--1.3 \\
    I3 & 0.4--0.7 & 0.2--0.4 & 0.4--0.8 \\
    \bottomrule
  \end{tabular}
\end{table}

Among the indicators in Table~\ref{tab:early}, PiP is the most accurate at every measurement age on all three measures
(Figure~\ref{fig:early} shows precision and the AUC for a future top-1\% paper; the table, lead times on all three). At
career age 3,
the top 10\% of a cohort by PiP contains 21\% future stars, twice the chance rate, against 18\% for the $h$-index; at
age 5 the shares are 30\% and 25\%, and at age 10, 49\% and 42\%. The AUC for a future top-1\% paper is 0.63 against
0.56 at age 3 and 0.70 against 0.66 at age 5.

Because every indicator improves with age, the gaps translate into time (Table~\ref{tab:early}): across the three
measures, PiP reaches the $h$-index's age-10 accuracy 1.6 to 1.9 years earlier (1.6 to 1.8 on the two measures for
future stars), at about age 8, and the citation count's 2.6 to 3.2 years earlier. Its lead over the article count, the
top-10\% count and I3 is smaller, at most 1.6 years. For the 711,986 authors eligible at age 5, PiP reaches
the $h$-index's age-10 accuracy 1.5 to 1.8 years earlier and the citation count's 2.3 to 2.9 years earlier.

\paragraph{Without the h-index condition.} Requiring an $h$-index of three by career age 3 keeps only 21\% of the
authors who have three articles by then, and it selects the population by one of the indicators being compared. We
therefore repeated it for all 1,455,891 authors with at least three articles at career age 3, with peers defined by the
same rule. PiP still leads the $h$-index at every measurement age: the top 10\% by PiP contains 24\% future stars at age
3 against 20\% for the $h$-index, and 31\% against 27\% at age 5. PiP reaches the $h$-index's age-10 accuracy 1.2 to 1.7
years earlier. It also leads the indicators of Table~\ref{tab:early} on precision and the AUC for a future top-1\% paper
at every age, while I3 matches or beats its AUC for future stars from age 6.

On the other two outcomes the ordering at each age matches that of Section~\ref{sec:prediction}: the article count
predicts the number of future top-10\% papers slightly better than PiP at the earliest ages, and the mean percentile
predicts future citation impact best. PiP's early-detection advantage lies in finding the scholars at the top, who
combine many papers with highly cited ones, and in anticipating exceptional papers. PiP-simple is about as accurate
here: within 0.007 of PiP at every age, in every population we tested, and within 0.3 years in lead time; under the
three-article rule it is slightly ahead on finding future stars from career age 6 or 7. In early detection the gain over
the $h$-index therefore comes mainly from what PiP and PiP-simple share: percentiles within the publication year and the
cutoff.
\subsection{Later recognition}
\label{sec:honors}

So far we have judged citation indicators by later citations; as a check, we ask whether the indicators anticipate an
outcome that is not a citation count: later scientific honors. From Wikidata (Section~\ref{sec:data}) we take 65 senior
honors for research: membership of national academies (such as the US National Academy of Sciences, the Royal Society
and the Leopoldina); fellowships of learned societies (AAAS, ACM, IEEE, APS and others); and major prizes and
fellowships (such as the Guggenheim, MacArthur and Gairdner awards). We exclude plain society memberships, state orders,
student scholarships and the citation-based Clarivate Citation Laureates. We connect honors to Semantic Scholar authors
in two ways: through Wikidata items with a Semantic Scholar author ID (41,272 items, honored or not), and through items
with a Google Scholar ID (12,434 honorees), linked to Semantic Scholar by title matching (Section~\ref{sec:robustness}).

We measure each indicator at career age 5 or 10 as a percentile within the start-year cohort (start years 1960--2000).
So that every counted honor follows the evaluation, the main analysis uses only dated honors: we drop people whose
first dated honor came at or before that age or whose honors are all undated. Most honors come decades later. This
leaves 919 future honorees at age 5 and 1,308 at age 10. The ROC AUC measures how well each indicator's ranking
separates future honorees from the others, among all eligible authors and within two already notable groups: people
with a Wikidata item, and our Google Scholar profiles matched to Semantic Scholar (mostly faculty).

\begin{table}[t]
  \centering
  \small
  \caption{ROC AUC for holding a senior honor. Columns 2--5: indicators measured at career age 5 or 10 in Semantic
  Scholar against dated honors received later, among all eligible authors (the comparison uses every honoree and a 10\%
  random sample of the others: 919 future honorees among 78,431 sampled authors at age 5, and 1,308 among 161,609 at
  age 10), among people with a Wikidata item, and among our Google Scholar profiles matched to
  Semantic Scholar (mostly faculty; GS: Google Scholar). Columns 6--7: current Google Scholar scores of the 15,890
  profiled scholars who meet the eligibility rule and belong to a start-year cohort of at least 30 profiles, against
  holding an honor and having an English Wikipedia article (agreement, not prediction). PiP's bootstrap intervals are
  about $\pm 0.015$ for all authors and $\pm 0.025$ within the smaller groups; differences between indicators are
  more precise because they are paired.}
  \label{tab:honors}
  \setlength{\tabcolsep}{4.5pt}
  \begin{tabular}{lcccccc}
    \toprule
    & \multicolumn{4}{c}{Predicting later honors (Semantic Scholar)} & \multicolumn{2}{c}{Google Scholar, now} \\
    \cmidrule(lr){2-5}\cmidrule(lr){6-7}
    & All, age 5 & All, age 10 & Wikidata, 10 & GS faculty, 10 & Honors & Wikipedia \\
    Future honorees / positives & 919 & 1,308 & 629 & 755 & 1,497 & 2,160 \\
    \midrule
    \textbf{PiP} ($\tau = 0.5$) & 0.751 & 0.827 & 0.641 & 0.648 & 0.712 & 0.716 \\
    PiP ($\tau = 0.75$) & \textbf{0.761} & \textbf{0.836} & \textbf{0.658} & \textbf{0.654} & 0.711 & \textbf{0.721} \\
    $h$-index & 0.709 & 0.790 & 0.618 & 0.637 & 0.708 & 0.704 \\
    Citations & 0.700 & 0.798 & 0.637 & 0.638 & \textbf{0.714} & \textbf{0.721} \\
    Articles & 0.694 & 0.783 & 0.579 & 0.628 & 0.677 & 0.675 \\
    Top-10\% count & 0.748 & 0.826 & 0.656 & 0.649 & 0.709 & 0.720 \\
    I3 & 0.722 & 0.799 & 0.607 & 0.639 & 0.696 & 0.692 \\
    Mean percentile & 0.531 & 0.572 & 0.574 & 0.541 & 0.611 & 0.611 \\
    \bottomrule
  \end{tabular}
\end{table}

Among all eligible authors, PiP at career age 5 reaches an AUC of 0.75, against 0.71 for the $h$-index and 0.70 for
citations; at age 10, 0.83 against 0.79 and 0.80 (Table~\ref{tab:honors}). Paired bootstrap intervals for PiP's lead
over the $h$-index exclude zero at both ages ($+0.042$ [0.031, 0.052] and $+0.037$ [0.031, 0.043]). PiP ties the
top-10\% count ($+0.001$ [$-0.005$, 0.007] at age 10), and the impact end of the dial beats the default: at
$\tau = 0.75$ the AUC is 0.76 and 0.84 ($+0.011$ [0.005, 0.017] and $+0.009$ [0.005, 0.013] over $\tau = 0.5$, with
paired intervals also above zero in both notable groups at age 10). Recognition rewards highly cited work more than
volume: the article count,
which leans most toward volume, does worst apart from the mean percentile.

Within the two notable groups, every indicator separates future honorees only weakly (AUC 0.54--0.66 at career age
10). There, PiP is level with the $h$-index at age 5 ($-0.005$ [$-0.019$, 0.011] among people with a Wikidata item and
$+0.001$ [$-0.011$, 0.013] among the Google Scholar faculty; not in the table) and ahead at age 10, by 0.022 [0.014,
0.030] and 0.011 [0.005, 0.018]. The top-10\% count and PiP at $\tau = 0.75$ or higher do best. Counting undated
honors as later (1,064 honorees at age 5 and 1,555 at age 10) changes little: PiP's AUC is 0.75 against 0.71 for the
$h$-index at age 5, and 0.82 against 0.78 at age 10.

A last check measures agreement rather than prediction (columns 6--7 of Table~\ref{tab:honors}). On the Google Scholar
profiles, where scores and honors are measured at the same time, the indicators are close: PiP is slightly ahead of the
$h$-index for holding an honor ($+0.004$ [0.001, 0.008]) and for having a Wikipedia article ($+0.012$ [0.009, 0.015]).
Total citations and the top-10\% count do as well for holding an honor (their paired differences from PiP include zero)
and slightly better for having a Wikipedia article (by 0.005 and 0.004).

The recognition test therefore supports PiP over the $h$-index with an outcome outside citation counts, most clearly for
detecting future honorees among all scholars and early in their careers. In populations closer to faculty evaluation,
the gain over the $h$-index is small, and the top-10\% count does as well. The test also shows that the default cutoff
is not best for every purpose: to anticipate recognition, an evaluator should move the dial toward impact. Above $\tau
= 0.6$ the raw score drifts across cohorts (Section~\ref{sec:tau}), so such scores should be read as cohort percentiles,
as they are here.

Several caveats apply. Wikidata covers honors unevenly across countries and fields, so an author without a recorded
honor is not a confirmed non-recipient; older cohorts have had more time to receive senior honors; and honors depend
on nominations, networks and opportunity as well as on the work. These processes can strengthen or weaken an
indicator's association with recorded honors, so the AUCs measure agreement with recorded recognition, not with merit.

\subsection{Fields}
\label{sec:fields}

The main analysis pools fields: it ranks each paper against all articles from its year, although fields cite at
different rates. Two questions follow: whether PiP's lead holds within each field, and whether pooling favors some
fields. For the first, we assign each author a main field, the most frequent among their articles by career age 10
(an article with several fields counts in full for each), and repeat the analyses by field. Fifteen fields have at
least 3,000 eligible authors with start years 1975--2000. In every one, PiP predicts better than the $h$-index on all
three outcomes (future top-10\% count / top-1\% AUC / future mean percentile): for example, in economics PiP reaches
0.39 / 0.74 / 0.27 against 0.31 / 0.66 / 0.17 for the $h$-index, and in mathematics 0.38 / 0.75 / 0.28 against
0.29 / 0.68 / 0.19.

\begin{figure}[t]
  \centering
  \includegraphics[width=0.72\textwidth]{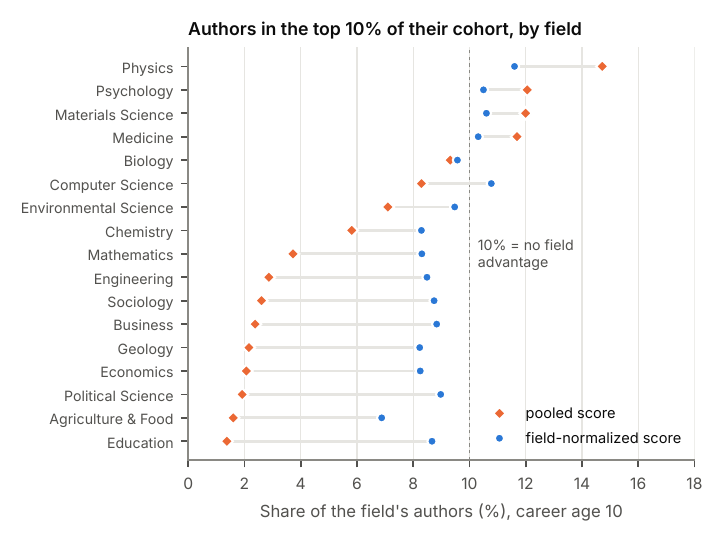}
  \caption{Share of each field's authors who are in the top 10\% of their start-year cohort at career age 10, with
  PiP computed from pooled percentiles and with the field-normalized version. Eligible authors with start years
  1975--2010; fields with at least 5,000 such authors.}
  \label{fig:fields}
\end{figure}

Pooled paper percentiles favor fields with high citation rates. A field-neutral score would put 10\% of each field's
authors in the top 10\% of their cohort; with pooled percentiles the share ranges from 1.4\% in education and 2.1\% in
economics to 14.7\% in physics (Figure~\ref{fig:fields}). We therefore also compute a field-normalized PiP. In this
version, each paper's percentile is computed within its publication year and the author's main field, and peers are
restricted to authors with the same main field and start year. A paper that does not carry the author's main field keeps
its pooled percentile. The rule also gives a coauthored paper different percentiles for coauthors with different main
fields, a choice that a rule based on the paper's own fields would avoid.

Among eligible authors at career age 10 with start years 1975--2010 (Figure~\ref{fig:fields}), the field-normalized
score puts every field's share of the cohort top 10\% between 6.9\% and 11.6\%. We have not separated the sources of the
remaining spread. Two candidates are that papers outside the author's main field keep their pooled percentile, and that
fields differ in how often peers tie, which also shapes the distribution of the score. Within each field, pooled and
field-normalized scores correlate at 0.91--0.99.

The two versions answer different questions: how a scholar's future work will rank among all articles of its year,
and how it will rank within its field. With outcomes measured on pooled percentiles, the pooled score predicts
best (0.427 / 0.772 / 0.321 against 0.414 / 0.770 / 0.298); with field-normalized outcomes, the field-normalized score
predicts best (0.407 / 0.772 / 0.316 against 0.405 / 0.769 / 0.300). Both remain far ahead of the $h$-index
(0.333 / 0.717 / 0.209 on field-normalized outcomes). For comparisons across fields we recommend the field-normalized
version; within a field the two are nearly interchangeable.

\subsection{Robustness}
\label{sec:robustness}

\paragraph{Self-citations.} Part of an indicator's predictive power could run through authors citing their own work. A
citation counts as a self-citation when the citing and cited papers share at least one author. Self-citations are
8.1\% of all citations; such rates depend strongly on the definition and the citation window \citep{aksnes2003macro}.
We removed them and reran the whole pipeline (percentiles, the $h$-index condition for eligibility, peer groups,
scores and outcomes; Table~\ref{tab:selfcites} in the appendix). PiP's predictions of the number of future top-10\%
papers and of a future top-1\% paper move by at most 0.003. The $h$-index loses slightly more (0.352 to 0.340 at age
10). The prediction of future citation impact drops more, and unevenly: by 0.018 to 0.030 for PiP and by 0.005 to
0.009 for the $h$-index. PiP's lead on this outcome therefore shrinks but remains (0.298 against 0.221 at career age
10), and PiP stays ahead of the $h$-index on all three outcomes in every setting. Removing self-citations changes
paper percentiles, eligibility, peer groups and outcomes at once, so the drop does not isolate the effect of authors
citing their own work.

\paragraph{Eligibility without the h-index condition.} The main rule requires an $h$-index of three, which selects the
population by citations and by one of the compared indicators and excludes far more authors in some fields than in
others (Section~\ref{sec:data}). Under the three-article rule, the same-career-age settings grow to 2.81, 2.54 and
2.00 million authors at career ages 5, 10 and 15 (from 1.17, 1.45 and 1.24 million). PiP still beats the $h$-index on
all three outcomes at every age, by 0.03 to 0.06, half to two thirds of the margin under the main rule; at career age
10 it reaches 0.420, 0.805 and 0.493 against 0.376, 0.770 and 0.443 (Table~\ref{tab:predictionall}). Every indicator
predicts future citation impact better than under the main rule (the mean percentile 0.519 against 0.394, the article
count 0.163 against 0.000 at career age 10), as expected when the population spans a wider range of records: the added
authors lack three articles with three citations each. PiP again has the highest AUC for a future top-1\% paper, and
no other indicator beats it on more than one outcome. Section~\ref{sec:early} reports early detection under the same
rule.

\paragraph{Author disambiguation.} Semantic Scholar's author records come from automatic disambiguation, which can split
one person across several records or merge different people. To check that the results are not artifacts of this
algorithm, we use the author-curated Google Scholar profiles. We matched the 16,218 profiles to Semantic Scholar through
their publications. A profile's work matches a Semantic Scholar paper on its normalized title (at least 25 characters)
and a publication year within one year. The profile's counterpart is the Semantic Scholar author record with the same
surname that holds the most matched works. We find a confident counterpart (at least five matched works and at least
30\% of the profile's titles) for 14,511 profiles (89\%). For the median person, 76\% of their Google Scholar works are
found in Semantic Scholar, and 90\% of those found sit under the counterpart. Splitting is common, though: 64\% of
people have at least two of their works under a second record with the same surname and first initial, and 42\% under a
third. The counterpart is smaller than the profile (median 67\% of the profile's cited works and 78\% of its
$h$-index), and its start year is later than the profile's for 47\% of people, because early work is missing or under
another record.\looseness=-1

Despite these differences, the two sources rank people similarly. For the 9,796 people who have scores in both sources
(Semantic Scholar start year 1995 or later, evaluated in 2025), the Spearman correlation between sources is 0.82 for the
$h$-index and 0.78 for PiP. Each source computes percentiles against its own population: all eligible Semantic Scholar
authors, and the Google Scholar profiles. For PiP, the agreement is lower for people split across three or more records
(0.73) than for people with at most one record (0.76), but not by much. These comparisons show that current rankings
agree closely with author-curated records, though record splitting can distort individual Semantic Scholar scores and
reconstructed career histories.

\paragraph{Survivorship.} The population includes every eligible author, not only those who stayed in research, and
the prediction results of Section~\ref{sec:prediction} count authors who stop publishing as having no future top
papers. This choice favors indicators that predict survival, that is, whether an author continues to publish. Measured
at career age 5, the indicators predict whether an author is still publishing at career ages 11--15 with an AUC of
0.68 for the article count, 0.62 for PiP, 0.58 for the $h$-index and 0.54 for citations. Continuing to publish is a
matter of volume, so the indicators that weigh volume most predict it best. PiP still leads the article count, the
best predictor of survival, on the top-1\% outcome and on future citation impact.
\section{Discussion}
\label{sec:discussion}

\paragraph{Why PiP improves on the $h$-index.} The $h$-index summarizes a record by one point; PiP uses every paper
above the cutoff, comparing the scholar with peers at every impact threshold. Three properties explain most of the
predictive gain. First, PiP credits the best papers in proportion to their rarity among peers: one top paper raises the
standing at the highest thresholds, where most peers have none. Second, PiP credits recent papers as soon as they pass
the cutoff among papers of the same age, whereas the $h$-index waits for enough raw citations. This credit matters most
early in a career, when most papers are recent, and likely explains much of PiP's lead in early detection
(Section~\ref{sec:early}). Third, PiP's scale does not depend on the start year: at the same career age, a given score
is about equally rare in every cohort (Section~\ref{sec:cohorts}), and in pools of mixed seniority the raw score
predicts almost as well as its cohort percentile (Section~\ref{sec:prediction}). The raw $h$-index instead favors longer
careers and later cohorts, which started when citation rates were higher.

Comparing PiP with PiP-simple shows what each part of PiP does. PiP-simple uses the same percentiles and cutoff, and so
shares the credit for recent papers, but compares the scholar with peers only once. Percentiles within the publication
year and the cutoff carry most of the gain over the $h$-index in predicting future top papers and in early detection,
where PiP-simple is about as early as PiP. The comparison at every threshold (the first property) adds the gain on
future citation impact and the chance of an exceptional paper.

\paragraph{Prediction and cumulative advantage.} Citations reward visibility as well as research quality: well-known
scholars attract more citations for comparable work \citep{merton1968matthew,petersen2014reputation}. For any
citation-based indicator, PiP and the $h$-index alike, part of the predictive power may therefore reflect reputation
feeding itself (cumulative advantage) rather than persistent research quality. The most mechanical channel of this
feedback is self-citation. Removing self-citations narrows PiP's lead on future citation impact but leaves it ahead of
the $h$-index on all three outcomes (Section~\ref{sec:robustness}). This check does not rule out the broader effect of
reputation; separating that effect from research quality would need variation in visibility unrelated to the work's
quality. Indicators can absorb cumulative advantage to different degrees, so it may also explain part of their
differences. The test against later honors (Section~\ref{sec:honors}) uses a non-citation outcome and gives the same
ordering of PiP and the $h$-index, although visibility also shapes honors.

\paragraph{Fields and windows.} The pooled score ranks each paper against all articles of its publication year,
whatever their field, and so favors high-citation fields; the field-normalized version removes most of that advantage.
Field normalization brings its own choices: the field scheme and the assignment of multi-field papers
\citep{zitt2005relativity,waltman2015field,waltman2019field} and, for a scholar, of a main field. Semantic Scholar's 23
fields are broad, and subfields with different citation habits share a reference set, so normalization removes only
differences between these broad fields.

Percentiles of recent papers are also noisier in slow-citing fields such as the humanities and law
(Section~\ref{sec:paperstability}), so in those fields PiP's credit for recent papers is least reliable and part of a
scholar's score rests on percentiles still settling. An evaluation there might count only papers older than two or
three years, at the cost of ignoring the most recent work.

\paragraph{Limitations.}
\begin{itemize}
  \item \emph{Disambiguation.} Semantic Scholar's automatic disambiguation can split one person's work across several
  author records or merge different people's work into one. Split records shorten careers and move start years,
  placing the person in a later cohort with a partial record. Evaluations of individuals should use a curated record
  such as a Google Scholar or ORCID profile.
  \item \emph{Whole counting.} Every coauthor receives full credit for every paper, whatever the number of authors or
  their roles. Compared with fractional counting, it favors authors in large teams. A fractional version would weight
  each paper by the author's share when counting papers above each threshold \citep{waltman2015field}; we have not
  tested it.
  \item \emph{The peer group.} Peers share a start year, the year of the first article. Authors who start late,
  interrupt their careers or move between research and other work are compared with a cohort whose careers ran
  differently. The eligibility rule (three articles and an $h$-index of three) excludes much of the earliest career
  stage and of slow-citing fields; under the three-article rule the main comparisons hold with smaller margins
  (Sections~\ref{sec:early} and~\ref{sec:robustness}). Neither rule admits a scholar with one top paper and nothing else
  (Section~\ref{sec:properties}); such a record can be scored against a cohort, but our tests do not cover it.
  \item \emph{Retrospective reconstruction.} Publications and citations are counted up to each evaluation year, but
  author records, publication types and fields come from the 2026 release; an evaluation at the time would have seen
  different records.
  \item \emph{Reference set.} Each paper's reference set depends on Semantic Scholar's publication types, which are
  incomplete; untyped records are excluded from the reference sets, the paper counts and the $h$-index.
\end{itemize}

\paragraph{Responsible use.} PiP is meant to inform comparisons among scholars at a similar career stage, alongside
reading the work itself and expert judgment, as the Leiden Manifesto \citep{hicks2015bibliometrics} and DORA
\citep{dora2012} recommend. Its cutoff should be chosen before an evaluation, not tuned to favor a candidate, and
comparisons across fields should use the field-normalized version. A predictive correlation of 0.4 leaves most of a
scholar's future unexplained.

\section{Conclusion}
\label{sec:conclusion}

PiP rates a scholar by the citation percentiles of their papers above a cutoff, comparing the scholar at every impact
threshold with peers who started in the same year. For 11.1 million Semantic Scholar authors, PiP's distribution changes
little across cohorts since 1960, and a scholar's PiP is more stable over a career than their $h$-index. PiP predicts
three outcomes better than the $h$-index in every setting and field we tested: the number of future top papers, the
chance of an exceptional paper and future citation impact. No other indicator built from the same paper percentiles
beats PiP on more than one of them. PiP reaches the $h$-index's age-10 accuracy in detecting future high performers 1.6
to 1.8 years earlier, and it better anticipates later academy memberships, fellowships and prizes. The results hold
without self-citations and when eligibility does not depend on citations, PiP rankings agree between Semantic Scholar
and author-curated Google Scholar profiles, and a field-normalized version removes most of the high-citation field
advantage.

A single parameter, the cutoff, moves PiP between quantity and impact. The construction needs only each scholar's
articles, with their publication years and citations by year; \url{https://www.pip-score.org} provides an interactive
demonstration of the metric on Google Scholar profiles, and the replication code computes it at scale in a database.
\section*{Author contributions}
Panos Ipeirotis: Conceptualization, Methodology, Software, Formal analysis, Data curation, Visualization, Supervision,
Writing -- original draft, Writing -- review \& editing. Sen Tian: Conceptualization, Methodology, Formal analysis,
Writing -- review \& editing.

\section*{Competing interests}
Panos Ipeirotis developed and maintains \url{https://www.pip-score.org}, a free, non-commercial website computing
scholar indicators from Google Scholar profiles. The authors have no other competing interests.

\section*{Funding information}
No specific funding was received for this work.

\section*{Data availability}
The replication package is archived on Zenodo \citep{ipeirotis2026replication}. It holds the SQL that builds every
table, the 28 aggregate result tables that reproduce every table and number in the paper, the Lean~4 and numerical
checks of the properties in Section~\ref{sec:pip}, the data behind every figure, the plotting scripts, and the scripts
and curated honors of Section~\ref{sec:honors}. Its \texttt{COMPUTATION.md} explains how the SQL evaluates the integral
that defines PiP. Semantic Scholar data are available under the ODC-BY license from the Semantic Scholar Datasets API
\citep{s2datasets2026}; Wikidata is available under CC0 \citep{wikidata2026}. For legal reasons (Google Scholar's terms
of service), raw per-article data from Google Scholar profiles cannot be made openly available; aggregated author-level
statistics for the profiles can be downloaded from \url{https://www.pip-score.org/data}, and the aggregate validation
outputs are included in the Zenodo archive. Because the intermediate article-level reconstruction and per-author score
tables ($\approx$527~GB) exceed Zenodo's per-record size limit, they are hosted in the publicly queryable BigQuery
datasets \texttt{scholar-version2.paper\_rp} and \texttt{scholar-version2.paper\_rp\_noself} (without self-citations),
which anyone signed in to Google Cloud can query (queries are billed to the reader's own project).

\bibliographystyle{qss-apa}
\bibliography{references}

\appendix
\section{Self-citations}
\label{app:selfcites}

\begin{table}[ht]
  \centering
  \small
  \caption{Prediction results with and without self-citations (with / without). Self-citations are removed
  throughout: in the paper percentiles, the peer groups, the scores and the outcomes. Raw scores, pooled over start
  years; for the year 2005, Table~\ref{tab:prediction} reports cohort percentiles, which differ slightly.}
  \label{tab:selfcites}
  \begin{tabular}{llccc}
    \toprule
    Setting & Indicator & Top-10\% count & Top-1\% AUC & Mean pct. \\
    \midrule
    Age 5 & PiP & .378 / .378 & .715 / .712 & .264 / .234 \\
          & $h$-index & .268 / .257 & .646 / .640 & .155 / .148 \\
    \addlinespace
    Age 10 & PiP & .427 / .427 & .772 / .771 & .321 / .298 \\
           & $h$-index & .352 / .340 & .719 / .714 & .230 / .221 \\
    \addlinespace
    Age 15 & PiP & .436 / .437 & .799 / .799 & .343 / .325 \\
           & $h$-index & .384 / .374 & .758 / .754 & .270 / .261 \\
    \addlinespace
    Year 2005 & PiP & .407 / .407 & .776 / .777 & .310 / .291 \\
              & $h$-index & .284 / .277 & .691 / .689 & .174 / .169 \\
    \bottomrule
  \end{tabular}
\end{table}

\clearpage
\section{Prediction results in all settings}
\label{app:prediction}

\begin{table}[ht]
  \centering
  \footnotesize
  \setlength{\tabcolsep}{4pt}
  \caption{Prediction of the next ten years of work in all five settings, and at the same career age under the
  three-article rule: Spearman correlation with the number of future top-10\% articles, ROC AUC for at least one future
  top-1\% article, and Spearman correlation with the mean percentile of future articles. Same-year settings include
  career ages 2--30. PiP is the raw score pooled over start years (Table~\ref{tab:prediction} reports cohort
  percentiles in the same-year settings); the $h$-index is given raw and as a cohort percentile.}
  \label{tab:predictionall}
  \begin{tabular}{lccccccccc}
    \toprule
    & \multicolumn{3}{c}{PiP ($\tau = 0.5$), raw} & \multicolumn{3}{c}{$h$-index, raw} & \multicolumn{3}{c}{$h$-index, cohort pct.} \\
    \cmidrule(lr){2-4}\cmidrule(lr){5-7}\cmidrule(lr){8-10}
    Setting & Top-10\% & Top-1\% & Mean pct. & Top-10\% & Top-1\% & Mean pct. & Top-10\% & Top-1\% & Mean pct. \\
    \midrule
    Age 5 & .378 & .715 & .264 & .268 & .646 & .155 & .256 & .647 & .178 \\
    Age 10 & .427 & .772 & .321 & .352 & .719 & .230 & .352 & .721 & .238 \\
    Age 15 & .436 & .799 & .343 & .384 & .758 & .270 & .385 & .760 & .272 \\
    Year 2000 & .381 & .761 & .288 & .269 & .682 & .152 & .328 & .721 & .232 \\
    Year 2005 & .407 & .776 & .310 & .284 & .691 & .174 & .363 & .737 & .241 \\
    \addlinespace
    \multicolumn{10}{l}{\emph{Three-article rule}} \\
    Age 5 & .394 & .761 & .512 & .338 & .719 & .452 & -- & -- & -- \\
    Age 10 & .420 & .805 & .493 & .376 & .770 & .443 & -- & -- & -- \\
    Age 15 & .423 & .830 & .475 & .393 & .802 & .432 & -- & -- & -- \\
    \bottomrule
  \end{tabular}
\end{table}

\clearpage
\section{Mathematical proofs and formal verification}
\label{app:proofs}
\newcommand{\Peers}{{\mathcal{P}\mkern2.5mu}}
\providecommand{\square}{\leavevmode\hbox{\vrule\vbox{\hrule width 0.5em\kern 0.5em\hrule}\vrule}}

This appendix states and proves the mathematical properties of the standing curve, the grid and integral forms of PiP,
the cohort-mean identity, the upward-rounding equivalence, and the order-statistics formula for exact integration in
Section~\ref{sec:pip} and \texttt{COMPUTATION.md}. Every proposition is formally verified in Lean~4 with Mathlib
(\texttt{paper/lean/} in the replication package; Table~\ref{tab:lean}).

Let $\Peers$ be a finite peer cohort of size $N = |\Peers| \ge 1$. Each author $b \in \Peers$ has a finite list of
article percentiles $(y_{b,1}, \dots, y_{b,n_b(0)})$ in $[0, 1)$, and at any threshold $\theta \in [0, 1]$ has qualifying
count $n_b(\theta) = |\{j : y_{b,j} \ge \theta\}| \in \mathbb{N}_0$. For a focal author $a \in \Peers$, the standing at
$\theta$ is $g_a(\theta) = N^{-1}\sum_{b \in \Peers} \mathbf{1}\{n_b(\theta) < n_a(\theta)\}$, and the continuous score
with cutoff $\tau \in [0, 1)$ is $\PiP_\tau(a) = (1-\tau)^{-1}\int_\tau^1 g_a(\theta)\,d\theta$. Given a non-empty finite
threshold grid $\mathcal{T} = \{t_1, \dots, t_K\} \subset [\tau, 1]$ of size $K \ge 1$ and counts $n_{b,k} = n_b(t_k)$,
the finite-grid score is
\begin{equation}
  S(a) = \frac{1}{K}\sum_{k=1}^{K} g_a(t_k)
  = \frac{1}{NK}\sum_{k=1}^{K}\sum_{b \in \Peers} \mathbf{1}\{n_{b,k} < n_{a,k}\}.
  \label{eq:gridscore}
\end{equation}

\paragraph{Proposition C.1 (Bounds, zero score below cutoff, and sawtooth step).}
\emph{Let $a \in \Peers$ ($N = |\Peers| \ge 1$).}
\begin{enumerate}
  \item[(i)] \emph{At every threshold $\theta$, $0 \le g_a(\theta) \le (N-1)/N < 1$, and thus both $S(a)$ and
  $\PiP_\tau(a)$ lie in $[0, (N-1)/N]$.}
  \item[(ii)] \emph{If $n_a(\theta) = 0$, then $g_a(\theta) = 0$. If every article of $a$ has $y_{a,j} < \tau$, then
  $S(a) = 0$ and $\PiP_\tau(a) = 0$.}
  \item[(iii)] \emph{If $\theta_1 \le \theta_2$ and $n_a(\theta_1) = n_a(\theta_2)$, then $g_a(\theta_1) \le
  g_a(\theta_2)$.}
\end{enumerate}

\noindent\emph{Proof.}
(i) The strict-comparison set $\{b \in \Peers : n_b(\theta) < n_a(\theta)\}$ has non-negative size and excludes $a$ by
irreflexivity of $<$, so its cardinality is at most $N - 1$. Dividing by $N > 0$ gives $0 \le g_a(\theta) \le (N-1)/N <
1$, and averaging over $k$ or integrating over $[\tau, 1]$ preserves both bounds.
(ii) Since $n_b(\theta) \ge 0$ for all $b \in \Peers$, no peer satisfies $n_b(\theta) < 0$, so $n_a(\theta) = 0$ implies
$g_a(\theta) = 0$. If $y_{a,j} < \tau$ for all $j$, then $n_a(\theta) = 0$ for every $\theta \ge \tau$, making every
summand of $S(a)$ and the integrand of $\PiP_\tau(a)$ zero.
(iii) For every $b \in \Peers$, $\theta_1 \le \theta_2$ implies $\{j : y_{b,j} \ge \theta_2\} \subseteq \{j : y_{b,j}
\ge \theta_1\}$ and thus $n_b(\theta_2) \le n_b(\theta_1)$. Whenever $n_b(\theta_1) < n_a(\theta_1) = n_a(\theta_2)$,
transitivity gives $n_b(\theta_2) < n_a(\theta_2)$, so $g_a(\theta_1) \le g_a(\theta_2)$. \hfill$\square$

\begin{table}[t]
  \centering
  \footnotesize
  \setlength{\tabcolsep}{4pt}
  \caption{Mapping between Appendix~\ref{app:proofs} and the machine-checked Lean~4 theorems in \texttt{paper/lean/PiP/}
  (Lean and Mathlib \texttt{v4.35.0}); \texttt{lake build} compiles all modules and checks their axioms via
  \texttt{\#print axioms}.}
  \label{tab:lean}
  \begin{tabular}{llp{8.1cm}}
    \toprule
    Result & Module (\texttt{PiP/}) & Principal Lean theorem names \tabularnewline
    \midrule
    Prop.\ C.1(i): Score bounds & \texttt{Basic.lean} &
      \raggedright\texttt{gridScore\_nonneg}, \texttt{gridScore\_le\_pred\_div\_card}, \texttt{gridScore\_lt\_one} \tabularnewline
    Prop.\ C.1(ii): Zero below cutoff & \texttt{Basic.lean} &
      \raggedright\texttt{standing\_eq\_zero\_of\_count\_eq\_zero}, \texttt{gridScore\_eq\_zero\_of\_all\_below\_cutoff} \tabularnewline
    Prop.\ C.1(iii): Sawtooth step & \texttt{Basic.lean} &
      \raggedright\texttt{standing\_mono\_threshold\_on\_const\_count} \tabularnewline
    Prop.\ C.2: Monotonicity & \texttt{Basic.lean} &
      \raggedright\texttt{standing\_mono\_focal}, \texttt{gridScore\_mono\_focal} \tabularnewline
    Prop.\ C.2(i)--(iii): Paper edits & \texttt{Basic.lean} &
      \raggedright\texttt{gridScore\_add\_paper\_le}, \texttt{gridScore\_raise\_paper\_le}, \texttt{gridScore\_add\_paper\_below\_cutoff\_eq} \tabularnewline
    Remark: Citation ties & \texttt{Basic.lean} &
      \raggedright\texttt{citation\_monotonicity\_counterexample} \tabularnewline
    Prop.\ C.3: Cohort mean & \texttt{CohortMean.lean} &
      \raggedright\texttt{cohort\_mean\_standing}, \texttt{cohort\_mean\_gridScore}, \texttt{cohort\_mean\_pip} \tabularnewline
    Prop.\ C.4: Strict rounding & \texttt{GridIntegral.lean} &
      \raggedright\texttt{roundUp\_ge\_iff\_ge\_grid}, \texttt{standingAt\_roundUp\_eq\_on\_cell}, \texttt{grid\_eq\_integral\_roundUp} \tabularnewline
    Prop.\ C.5: Exact integration & \texttt{ExactIntegral.lean} &
      \raggedright\texttt{countAboveList\_lt\_iff\_kthPercentile\_lt}, \texttt{integral\_standingAt\_const\_eq\_H\_sub\_H}, \texttt{H\_eq\_asof\_formula} \tabularnewline
    \bottomrule
  \end{tabular}
\end{table}

\paragraph{Proposition C.2 (Conditional monotonicity with peer-group update).}
\emph{Let $a \in \Peers$, and let $n, n' : \Peers \times \{1, \dots, K\} \to \mathbb{N}_0$ be two qualifying-count
profiles such that $n'_{b,k} = n_{b,k}$ for every peer $b \ne a$ and $n'_{a,k} \ge n_{a,k}$ for every threshold $k \in
\{1, \dots, K\}$. Recomputing standings within the updated cohort $\Peers$ (with $a$ included in $\Peers$ under both
profiles) gives $g'_a(t_k) \ge g_a(t_k)$ for all $k$ and $S'(a) \ge S(a)$. Thus, with peers' article percentiles
and the focal author's other article percentiles held fixed:}
\begin{enumerate}
  \item[(i)] \emph{adding an article with assigned percentile $y$ to author $a$ cannot lower $S(a)$ or $\PiP_\tau(a)$;}
  \item[(ii)] \emph{raising an article's assigned percentile from $y$ to $y' \ge y$ cannot lower $S(a)$ or
  $\PiP_\tau(a)$; and}
  \item[(iii)] \emph{adding an article with assigned percentile $y < \tau$ leaves $S(a)$ and $\PiP_\tau(a)$ unchanged.}
\end{enumerate}

\noindent\emph{Proof.}
Fix $k$ and any peer $b \in \Peers$ with $n_{b,k} < n_{a,k}$. Since $n_{a,k} < n_{a,k}$ is false, $b \ne a$, so
$n'_{b,k} = n_{b,k} < n_{a,k} \le n'_{a,k}$. Thus $\{b \in \Peers : n_{b,k} < n_{a,k}\} \subseteq \{b \in \Peers :
n'_{b,k} < n'_{a,k}\}$, which yields $g_a(t_k) \le g'_a(t_k)$ for each $k$ and $S(a) \le S'(a)$ upon averaging. Under
(i), appending $y$ increases $n_a(\theta)$ by $\mathbf{1}\{y \ge \theta\} \ge 0$ while leaving $n_b(\theta)$ unchanged
for $b \ne a$; under (ii), replacing $y$ by $y' \ge y$ satisfies $\mathbf{1}\{y \ge \theta\} \le \mathbf{1}\{y' \ge
\theta\}$ at every $\theta$; under (iii), $y < \tau \le \theta$ implies $\mathbf{1}\{y \ge \theta\} = 0$ for all
$\theta \in [\tau, 1]$, so every count $n_b(\theta)$ on $[\tau, 1]$ is unchanged. \hfill$\square$

\paragraph{Remark (Citation-monotonicity counterexample).}
Proposition~C.2 holds fixed the assigned percentiles of other articles. When a paper instead gains raw citations and
same-year percentiles are recomputed via $y_i = |\{\ell : c_\ell < c_i\}|/M_{\text{year}}$, a tie with another of the
focal author's papers lowers that second paper's percentile and can reduce PiP. Consider a publication year with four
papers and citation counts $c = (0, 0, 1, 2)$, giving strict percentiles $(0, 0, 1/2, 3/4)$. Let $\Peers = \{a, b\}$,
where $a$ authored the third and fourth papers (percentiles $1/2$ and $3/4$) and $b$ authored the first two (percentile
$0$). On $(1/2, 3/4]$, $n_a(\theta) = 1 > 0 = n_b(\theta)$, so $g_a(\theta) = 1/2$ and $\PiP_{1/2}(a) = 1/4$. When $a$'s
third paper gains one citation, $c' = (0, 0, 2, 2)$ and the recomputed percentiles become $(0, 0, 1/2, 1/2)$; on $(1/2,
3/4]$, $n_a(\theta)$ drops from $1$ to $0$, reducing $g_a(\theta)$ from $1/2$ to $0$ and $\PiP_{1/2}(a)$ from $1/4$ to $0$.

\paragraph{Proposition C.3 (Cohort-mean tie identity).}
\emph{Let $p_{k,m} = N^{-1}|\{a \in \Peers : n_{a,k} = m\}|$ and $p_{\theta,m} = N^{-1}|\{a \in \Peers : n_a(\theta) =
m\}|$ denote the share of peers in $\Peers$ with qualifying count $m \in \mathbb{N}_0$. Then}
\begin{equation}
  \frac{1}{N}\sum_{a \in \Peers} S(a) = \frac{1}{2K}\sum_{k=1}^{K}\Bigl(1 - \sum_{m \ge 0} p_{k,m}^2\Bigr) < \frac{1}{2},
  \qquad
  \frac{1}{N}\sum_{a \in \Peers} \PiP_\tau(a)
  = \frac{1}{2(1-\tau)}\int_\tau^1 \Bigl(1 - \sum_{m \ge 0} p_{\theta,m}^2\Bigr)\,d\theta .
  \label{eq:gridmean}
\end{equation}

\noindent\emph{Proof.}
Fix $k$ and partition the $N^2$ ordered pairs $(a, b) \in \Peers \times \Peers$ by trichotomy into $W_k = \{(a,b) :
n_{b,k} < n_{a,k}\}$, $L_k = \{(a,b) : n_{a,k} < n_{b,k}\}$, and $T_k = \{(a,b) : n_{a,k} = n_{b,k}\}$. The involution
$(a, b) \mapsto (b, a)$ is a bijection between $W_k$ and $L_k$, so $|W_k| = |L_k|$, while grouping $T_k$ by the common
value $m = n_{a,k} = n_{b,k}$ gives $|T_k| = \sum_{m \ge 0} |\{a \in \Peers : n_{a,k} = m\}|^2 = N^2 \sum_{m \ge 0}
p_{k,m}^2$. Because $|W_k| + |L_k| + |T_k| = 2|W_k| + |T_k| = N^2$ and $\sum_{a \in \Peers} g_a(t_k) = |W_k|/N$,
dividing by $2N^2$ gives $N^{-1}\sum_{a \in \Peers} g_a(t_k) = \frac{1}{2}(1 - \sum_{m \ge 0} p_{k,m}^2)$. Since $N \ge
1$, at least one count $m$ has $p_{k,m} > 0$, so $\sum_{m \ge 0} p_{k,m}^2 > 0$ and the mean is strictly below $1/2$.
Averaging over $k = 1, \dots, K$ gives the first identity in Eq.~\ref{eq:gridmean}, and integrating over $\theta \in
[\tau, 1]$ gives the second (Section~\ref{sec:definition}). Five conventions enter: every author is evaluated against
$\Peers$; the focal author remains in $\Peers$ (contributing $(a,a) \in T_k$); comparisons are strict; peers with
$n_{b,k} = 0$ stay in $\Peers$; and the pair interpretation samples with replacement. \hfill$\square$

\paragraph{Proposition C.4 (Grid-to-integral equivalence under strict upward rounding).}
\emph{For an integer grid size $M \ge 1$ and cutoff index $0 \le q < M$ with $\tau = q/M$, define $R_M(y) = (\lfloor M y
\rfloor + 1)/M$ and let $g_a^{R_M}(\theta)$ be author $a$'s standing at $\theta$ after replacing every article percentile
$y$ in $\Peers$ by $R_M(y)$. For $0 \le k < M$ and $\theta \in (k/M, (k+1)/M]$, $g_a^{R_M}(\theta) = g_a(k/M)$ and}
\begin{equation}
  \frac{1}{M - q}\sum_{k=q}^{M-1} g_a(k/M) = \frac{1}{1-\tau}\int_\tau^1 g_a^{R_M}(\theta)\, d\theta .
  \label{eq:gridequiv}
\end{equation}

\noindent\emph{Proof.}
Fix $k \in \{0, \dots, M-1\}$ and $\theta \in (k/M, (k+1)/M]$. Because $M\theta \in (k, k+1]$ and $\lfloor M y \rfloor$
is an integer, $\lfloor M y \rfloor + 1 \ge M\theta$ holds if and only if $\lfloor M y \rfloor + 1 \ge k + 1$, which is
equivalent to $\lfloor M y \rfloor \ge k$ and hence to $y \ge k/M$. Dividing by $M > 0$ shows that $R_M(y) \ge \theta
\iff y \ge k/M$ for every article of every peer $b \in \Peers$. Thus $n_b^{R_M}(\theta) = n_b(k/M)$ for all $b \in
\Peers$, so $g_a^{R_M}(\theta) = g_a(k/M)$ on $(k/M, (k+1)/M]$ (length $1/M$), with cell integral $\int_{k/M}^{(k+1)/M}
g_a^{R_M}(\theta)\, d\theta = M^{-1} g_a(k/M)$. Summing $k = q, \dots, M-1$ from $\tau = q/M$ to $1$ and dividing by
$1 - \tau = (M-q)/M$ gives Eq.~\ref{eq:gridequiv}; in particular, an article at $y = \tau$ has $R_M(\tau) = (q+1)/M$ and
qualifies throughout $(\tau, (q+1)/M]$. \hfill$\square$

\paragraph{Proposition C.5 (Exact integration via order statistics).}
\emph{For each peer $b \in \Peers$, sort $b$'s article percentiles in non-increasing order $y_{b,(1)} \ge y_{b,(2)} \ge
\cdots \ge y_{b,(n_b(0))}$ and set $y_{b,(k)} = 0$ for $k > n_b(0)$. Then for every integer $k \ge 1$ and threshold
$\theta > 0$, $n_b(\theta) < k$ if and only if $y_{b,(k)} < \theta$. Therefore, on any interval $(x_0, x_1] \subseteq
[\tau, 1]$ with $0 \le x_0 \le x_1$ on which the focal author has constant count $n_a(\theta) = k \ge 1$,}
\begin{equation}
  \int_{x_0}^{x_1} g_a(\theta)\, d\theta = H_k(x_1) - H_k(x_0),
  \qquad
  H_k(x) = \frac{1}{N}\sum_{b \in \Peers} \max\!\bigl(x - y_{b,(k)},\, 0\bigr)
  = \frac{c_k(x)\, x - s_k(x)}{N},
  \label{eq:exactint}
\end{equation}
\emph{where $c_k(x) = |\{b \in \Peers : y_{b,(k)} \le x\}|$ and $s_k(x) = \sum_{b \in \Peers : y_{b,(k)} \le x}
y_{b,(k)}$.}

\noindent\emph{Proof.}
Since $b$'s percentiles are non-increasing in $[0, 1)$, at least $k$ entries satisfy $y_{b,j} \ge \theta > 0$ if and only
if $k \le n_b(0)$ and the $k$-th entry satisfies $y_{b,(k)} \ge \theta$; since $y_{b,(k)} = 0 < \theta$ when $k >
n_b(0)$, negation gives $n_b(\theta) < k \iff y_{b,(k)} < \theta$. Thus, for $\theta \in (x_0, x_1]$ where $n_a(\theta) =
k$, $g_a(\theta) = N^{-1}\sum_{b \in \Peers} \mathbf{1}\{y_{b,(k)} < \theta\}$. For each peer $b$, $\int_{x_0}^{x_1}
\mathbf{1}\{y_{b,(k)} < \theta\}\, d\theta = \max(x_1 - \max(x_0, y_{b,(k)}), 0) = \max(x_1 - y_{b,(k)}, 0) - \max(x_0 -
y_{b,(k)}, 0)$ whenever $x_0 \le x_1$. Averaging over $b \in \Peers$ gives $H_k(x_1) - H_k(x_0)$. Finally, partitioning
$\Peers$ into $\{b : y_{b,(k)} \le x\}$ (where $\max(x - y_{b,(k)}, 0) = x - y_{b,(k)}$) and $\{b : y_{b,(k)} > x\}$
(where $\max(x - y_{b,(k)}, 0) = 0$) gives $N H_k(x) = c_k(x)\, x - s_k(x)$, the running count-and-sum formula evaluated
by \texttt{sql/25\_pip\_exact.sql}. \hfill$\square$

\end{document}